\documentclass{article}
\usepackage[utf8]{inputenc}

\usepackage{xcolor}
\definecolor{blueviolet}{rgb}{0.2, 0.2, 0.6}
\definecolor{webgreen}{rgb}{0,.5,0}
\definecolor{webbrown}{rgb}{.6,0,0}
\usepackage[pdftex,
  bookmarks=false,
  colorlinks=true, 
  urlcolor=webbrown,
  linkcolor=blueviolet, 
  citecolor=webgreen,
  pdfstartpage=1,
  pdfstartview={FitH},  
  bookmarksopen=false,
  pagebackref
  ]{hyperref}
\renewcommand*{\backref}[1]{}
\renewcommand*{\backrefalt}[4]{%
    \ifcase #1%
          \or Cited on page~#2.%
          \else Cited on pages~#2.%
    \fi%
    }
\usepackage{amsmath}
\usepackage{amssymb}
\usepackage{amsfonts}
\usepackage{amsthm}
\usepackage{cleveref}
\allowdisplaybreaks 

\usepackage{enumerate}
\usepackage{comment}
\usepackage{xpatch}
\usepackage{graphicx}
\usepackage{tabularx}
\usepackage{braket}
\usepackage{physics}
\usepackage{amsthm}
\usepackage{tikz}
\usepackage{commath}
\usepackage{mathtools}
\usepackage{qcircuit}
\usepackage{algorithm}
\usepackage{algpseudocodex}[indLines = true,italicComments = false]
\usepackage{tabto}
\usepackage{pgf-umlsd}
\usepackage{bm}
\usepackage{bbm}
\usepackage{multirow, multicol}
\usepackage{float}
\usepackage{pdfpages}
\usepackage{caption}
\usepackage{subcaption}
\usepackage{makecell}
\usepackage{cite}
\usepackage{tablefootnote}

\DeclarePairedDelimiter\ryanabs{\lvert}{\rvert}

\DeclareMathOperator*{\argmax}{arg\,max}

\let\originalleft\left
\let\originalright\right
\renewcommand{\left}{\mathopen{}\mathclose\bgroup\originalleft}
\renewcommand{\right}{\aftergroup\egroup\originalright}

\newcommand{\mc}[1]{\mathcal{#1}}
\newcommand{\mbb}[1]{\mathbb{#1}}
\newcommand{\mrm}[1]{\mathrm{#1}}

\newcommand{\dtr}{\mathrm{d_{tr}}}
\newcommand{\dtv}{\mathrm{d_{TV}}}

\newcommand{\divchi}{\mathrm{d}_{\chi^2}}
\newcommand{\qdivchi}{\mathrm{D}_{\chi^2}}

\newcommand{\bigo}{\mathcal{O}}

\newcommand{\mmstate}{\frac{\mathbbm{1}}{d}}

\newcommand{\diag}{\mathrm{diag}}
\newcommand{\ssing}{S_{\mathrm{sing}}}
\newcommand{\stail}{S_{\mathrm{tail}}}
\newcommand{\smany}{S_{\mathrm{many}}}

\newcommand{\imax}{i_{\max}}
\newcommand{\srem}{S_{\mathrm{periph}}}
\newcommand{\bu}{\boldsymbol{u}}
\newcommand{\bv}{\boldsymbol{v}}
\newcommand{\bx}{\boldsymbol{x}}
\newcommand{\bfu}{\boldsymbol{U}}
\newcommand{\bfv}{\boldsymbol{V}}
\newcommand{\bftheta}{\boldsymbol{\theta}}
\newcommand{\bfDelta}{\boldsymbol{\Delta}}
\newcommand{\HS}{2}
\newcommand{\oper}{\mathrm{op}}
\newcommand{\eps}{\epsilon}

\newtheorem{theorem}{Theorem}[section]

\newtheorem{lemma}[theorem]{Lemma}
\newtheorem{corollary}[theorem]{Corollary}

\newtheorem{fact}[theorem]{Fact}

\theoremstyle{definition}
\newtheorem{definition}[theorem]{Definition}

\numberwithin{equation}{section}

\makeatletter
\newif\if@comments
\newcommand{\chirag}[1]{\if@comments\textcolor{blue}{[CW: #1]}\fi}
\newcommand{\ryan}[1]{\if@comments\textcolor{pink}{[RO: #1]}\fi}

\newcommand\enablecomments{\@commentstrue}
\newcommand\disablecomments{\@commentsfalse}
\enablecomments

\makeatother
\usepackage[a4paper, left=1in, right=1in, top=1in, bottom=1in]{geometry}

\title{Instance-Optimal Quantum State Certification \\with Entangled Measurements}

\author{
Ryan O'Donnell\thanks{Carnegie Mellon University Computer Science Department. Email: \href{mailto:odonnell@cs.cmu.edu}{odonnell@cs.cmu.edu}.  Supported in part by a grant from Google Quantum AI.}
\and
Chirag Wadhwa\thanks{School of Informatics, University of Edinburgh. Email: \href{mailto:chirag.wadhwa@ed.ac.uk}{chirag.wadhwa@ed.ac.uk}.} 
}
\date{}
\begin{document}

\maketitle

\disablecomments
\begin{abstract}

We consider the task of quantum state certification: given a description of a hypothesis state~$\sigma$ and multiple copies of an unknown state~$\rho$, a tester aims to determine whether the two states are equal or $\epsilon$-far in trace distance. It is known that~$\Theta(d/\epsilon^2)$ copies of~$\rho$ are necessary and sufficient for this task, assuming the tester can make entangled measurements over all copies~\cite{childs2007weak,o2015quantum,buadescu2019quantum}. However, these bounds are for a worst-case~$\sigma$, and it is not known what the optimal copy complexity is for this problem on an \emph{instance-by-instance} basis. While such instance-optimal bounds have previously been shown for quantum state certification when the tester is limited to  measurements unentangled across copies~\cite{chen2022toward,chen2022tightStateCertification}, they remained open when testers are unrestricted in the kind of measurements they can perform.

We address this open question by proving nearly instance-optimal bounds for quantum state certification when the tester can perform fully entangled measurements. Analogously to the unentangled setting, we show that the optimal copy complexity for certifying~$\sigma$ is given by the worst-case complexity times the fidelity between~$\sigma$ and the maximally mixed state. 
We prove our lower bounds using a novel quantum analogue of the Ingster--Suslina method, which is likely to be of independent interest. This method also allows us to recover the~$\Omega(d/\epsilon^2)$ lower bound for mixedness testing~\cite{o2015quantum}, i.e., certification of the maximally mixed state, with a surprisingly simple proof. 
\end{abstract}

\newpage

\tableofcontents

\newpage
\section{Introduction}
\label{sec:intro}
We revisit the problem of state certification for mixed quantum states. Here, given the description of a hypothesis state $\sigma$ and multiple copies of an unknown state $\rho$, a tester aims to determine (with high probability) whether $\rho = \sigma$ or $\|\rho - \sigma\|_1 \geq \epsilon$, promised that one of the two cases holds. This problem is analogous to the classical problem of \emph{identity testing} of distributions (or \emph{goodness of fit} testing). Quantum state certification is also relevant in practical settings, with potential applications in benchmarking quantum devices and verifying heuristic quantum algorithms.

A primary direction of study for this problem is the \emph{copy complexity}, i.e., how many copies of~$\rho$ are used. The optimal copy complexity for this task is known to be $\Theta(d/\epsilon^2)$ \cite{childs2007weak,o2015quantum,buadescu2019quantum}. However, these bounds are worst-case in nature, with the lower bound being shown for the specific case of \emph{mixedness testing}, where $\sigma = \mmstate$. Such worst-case bounds do not rule out the possibility that the problem is significantly easier for other hypotheses. Indeed, for the least mixed states --- i.e., those $\sigma$ that are pure (rank~$1$) --- a straightforward algorithm can be used for state certification with just $\bigo(1/\epsilon^2)$ copies (see e.g.,~\cite[Section 4.1.1]{montanaro2016survey}). 

To characterize the landscape between these two extremes, we consider the \emph{instance-optimal} setting. Here, one aims to show optimal bounds depending on the specific choice of the hypothesis state~$\sigma$. Nearly instance-optimal bounds have already been shown by Chen et~al.~\cite{chen2022toward,chen2022tightStateCertification} in the setting where testers are restricted to \emph{unentangled measurements} over individual copies of~$\rho$.
In particular, let $\underline{\sigma}$ and $\overline{\sigma}$ be quantum states obtained by zeroing out suitably chosen eigenvalues of $\sigma$ adding up to at most $\bigo(\epsilon)$ and $\bigo(\epsilon^2)$ respectively and then normalizing. Then, Chen et al.~showed that the optimal copy complexity $n$ satisfies
\begin{equation}
\label{eq:intro-unentangled-bounds}
    \Tilde{\Omega} \left(\frac{d \cdot \sqrt{\mathsf{rank}(\underline{\sigma})}}{\epsilon^2} \cdot F(\underline{\sigma}, \mathbbm{1}/d)\right) \leq n \leq \Tilde{\bigo} \left(\frac{d \cdot \sqrt{\mathsf{rank}(\overline{\sigma})}}{\epsilon^2} \cdot F(\overline{\sigma}, \mathbbm{1}/d)\right).
\end{equation}

However, when testers are unrestricted and allowed to perform entangled measurements over all copies of $\rho$, only the aforementioned worst-case bounds are known. This raises our central question:

\begin{center}
    \emph{
    How does the copy complexity of quantum state certification depend on the hypothesis state\\ when testers can perform general (fully entangled) measurements?
    }
\end{center}

\subsection{Our Results}
Before stating our results, we formally define the task of quantum state certification. We will follow this definition throughout the rest of this paper.

\begin{definition}[Quantum State Certification with Entangled Measurements]
\label{def:certification}
    Let $\sigma \in \mbb{C}^{d \times d}$ be a density matrix whose explicit description is given. Let $\rho \in \mathbb{C}^{d \times d}$ be an unknown quantum state, with the promise that either $\rho = \sigma$ or $\|\rho - \sigma\|_1 \geq \epsilon$. Then, the task of $\eps$-\emph{certifying}~$\sigma$ refers to distinguishing between these two cases with probability at least~$\tfrac23$ using measurements that may be fully entangled over all copies of~$\rho$.\footnote{As usual, $\frac23$ can be boosted to any $1-\delta$ at the expense of $\bigo(\mathrm{log}(1/\delta))$ times more copies.
    }
\end{definition}

 We now answer our central question by proving nearly instance-optimal bounds, encapsulated in the following theorem.

\begin{theorem}[Informal, see \Cref{thm:lower-main,thm:upper-bound}]
\label{thm:main-result}
     Fix $0 < \eps \leq 1/24$. Given a quantum state $\sigma \in \mathbb{C}^{d \times d}$, let $\underline{\sigma}$ and $\overline{\sigma}$ denote states obtained by zeroing out some suitably chosen eigenvalues of $\sigma$ adding up to $\bigo(\epsilon)$ and $\bigo(\epsilon^2)$ respectively and then normalizing\footnote{While we reuse $\underline{\sigma},\overline{\sigma}$ for notational convenience, our construction of $\underline{\sigma}$ differs slightly from that in \Cref{eq:intro-unentangled-bounds}. We elaborate on this in \Cref{sec:bucketing-lower}.
     }. Then, the copy complexity~$n$ of $\eps$-certifying~$\sigma$ 
     satisfies
    \begin{equation}
        \Tilde{\Omega}\left(
            \frac{d \cdot F(\underline{\sigma}, \mathbbm{1}/d)}{\epsilon^2} 
        \right)
        \leq 
        n
        \leq
        \Tilde{\bigo}\left(
            \frac{d \cdot F(\overline{\sigma}, \mathbbm{1}/d)}{\epsilon^2}
        \right),
    \end{equation}
    where $F$ denotes the fidelity of two quantum states.
\end{theorem}
First, comparing the above theorem to the unentangled-measurement bounds in \Cref{eq:intro-unentangled-bounds}, we see that with entangled measurements, the copy complexity of state certification is essentially improved by factors of $\sqrt{\mathsf{rank}(\underline{\sigma})}$ and $\sqrt{\mathsf{rank}(\overline{\sigma})}$ in the lower and upper bounds respectively. 

 We also note that our bounds are optimal (up to~$\mathrm{log} (d/\epsilon)$ factors) in both extremes discussed previously. When~$\sigma$ is a pure state, none of the eigenvalues is removed, and so $n = \Tilde{\Theta}(1/\epsilon^2)$. At the other extreme, when $\sigma = \mmstate$, the mass removed does not significantly alter the fidelity, giving us $n = \Tilde{\Theta}(d/\epsilon^2)$ and recovering the optimal lower bound up to log factors. In fact, we also apply our new lower bound techniques directly to mixedness testing and recover the lower bound without any loss in log factors, as stated in the following theorem.
\begin{theorem}
\label{thm:intro-mixedness}
    Let $0 < \epsilon < 1$. Any algorithm for mixedness testing with fully entangled measurements must use $\Omega(d/\epsilon^2)$ copies.    
\end{theorem}

Unlike the original proof of \cite{o2015quantum}, which uses sophisticated representation-theoretic tools, our proof only requires some straightforward calculations. Further, the original proof requires that the hypothesis state be \emph{exactly} maximally mixed. On the other hand, our techniques are robust and extend essentially automatically to states that are only \emph{nearly maximally mixed}. We state a more general version of this result in the following theorem:

\begin{theorem}
\label{thm:intro-almost-mixedness}
Let $0 < \epsilon \leq 1/2$. Let $\sigma \in \mathbb{C}^{d \times d}$ be a quantum state with all eigenvalues at least $2\eps/d$. Then, the copy complexity of $\eps$-certifying~$\sigma$ is at least
 \begin{equation}
        \Omega\left(\frac{d^{5/2}}{\|\sigma^{-1}\|_{\HS} \cdot \eps^2}\right).
\end{equation}
\end{theorem}

In particular, when the above bound is applied to a hypothesis state with all eigenvalues $\Omega(1/d)$, we again obtain an $\Omega(d/\epsilon^2)$ bound. While it may not be immediately clear why such states are relevant, we remark that they occur naturally while proving instance-optimal lower bounds, making the above theorem an important step towards our lower bound in \Cref{thm:main-result}. We elaborate on this in \Cref{sec:techniques}.

\subsection{Related Work}
\label{sec:related}
\paragraph{Instance-Optimal Distribution Testing.}
In the classical case of identity testing for probability distributions, instance-optimal bounds were first shown by \cite{valiant2017automatic}. Here, given the description of a distribution $q$, and samples from an unknown distribution $p$, one wishes to test whether the distributions are identical or $\epsilon$-far. Later, \cite{diakonikolas2016new,blais2019distribution} also proved such instance-optimal bounds using different sets of techniques. \cite{diakonikolas2016new} further showed instance-optimal bounds for the harder problem of \emph{closeness testing}, where both distributions are unknown. We refer to \cite{goldreich2017introduction,canonne2020survey,canonne2022topics} and references therein for detailed reviews of the broader field of distribution testing.

\paragraph{Mixedness Testing.}
For mixedness testing with entangled measurements, \cite{childs2007weak} showed that distinguishing the maximally mixed $d$-dimensional state from one that is maximally mixed on a random $d/2$-dimensional subspace requires $\Omega(d)$ copies.  For the general mixedness testing task, 
\cite{o2015quantum} showed that $\Theta(d/\epsilon^2)$ copies are necessary and sufficient. When testers are restricted to unentangled nonadaptive measurements, \cite{bubeck2020entanglement} showed that the optimal copy complexity is $\Theta(d^{3/2}/\epsilon^2)$. \cite{bubeck2020entanglement} also showed that $\Omega(d^{4/3}/\epsilon^2)$ copies are necessary even when the tester can adaptively choose measurements based on prior outcomes. \cite{chen2022tightStateCertification} later improved the lower bound for adaptive measurements to $\Omega(d^{3/2}/\epsilon^2)$, thus showing that $\Theta(d^{3/2}/\epsilon^2)$ copies are optimal for mixedness testing with unentangled measurements. When testers can perform entangled measurements on only a few copies of the unknown state at a time, \cite{chen2021hierarchy} showed an $\Omega(d^{4/3}/\epsilon^2)$ lower bound for this problem. Proving tight bounds for mixedness testing in such a setting remains an open problem.

\paragraph{Quantum State Certification.}
For certification with entangled measurements on general mixed states, \cite{buadescu2019quantum} gave an algorithm using $\bigo(d/\epsilon^2)$ copies. In fact, their algorithm also succeeds at closeness testing, i.e., when both $\rho$ and $\sigma$ are unknown. Paired with the lower bound of \cite{o2015quantum} for mixedness testing, this provides a tight worst-case characterization of state certification with entangled measurements. In the unentangled nonadaptive case, \cite{chen2022toward} presented a state certification algorithm using $\bigo(d^{3/2}/\epsilon^2)$ copies, matching the lower bounds of \cite{bubeck2020entanglement,chen2022tightStateCertification} for mixedness testing with unentangled measurements. \cite{chen2022toward} also proved instance-optimal bounds for state certification with unentangled, nonadaptive measurements, 
and \cite{chen2022tightStateCertification} later showed that the same complexity is optimal even when testers can perform adaptively chosen measurements. We also mention the problem of quantum state certification for pure $n$-qubit states when it is required to make unentangled measurements across \emph{qubits}.  In this case, a recent algorithm~\cite{gupta2025few} solves the problem with copy complexity~$O(n/\eps^2)$, following on from a work~\cite{huang2024certifying} that succeeded on a large class of pure states.

\medskip

Mixedness testing and quantum state certification are both \emph{quantum property testing} problems. A flurry of recent work has considered testing for various other properties of quantum states \cite {chen2022exponential,anshu2022distributed,chen2024optimalTradeoffsShadowTomography,gong2024sample} and quantum dynamics \cite{chen2022exponential,fawzi2023quantum,chen2023unitarity,bluhm2024hamiltonian}. For a more complete survey of quantum property testing, we refer to \cite{montanaro2016survey}.
\section{Technical Overview}
\label{sec:techniques}
Throughout, we will assume that $\sigma = \diag(\lambda_1, \dots, \lambda_d)$ is a diagonal density matrix. This is without loss of generality, as the tester is given the explicit description of $\sigma$ and can always change basis without using any additional copies. 

Recall that we stated \Cref{thm:main-result} in terms of $F(\underline{\sigma}, \mathbbm{1}/d)$ and $F(\overline{\sigma}, \mathbbm{1}/d)$. This interpretation in terms of fidelity arises from bounds that are directly comparable to the classical bounds of \cite{valiant2017automatic}. In the classical case, when one wishes to test identity to a distribution $q$, the sample complexity is characterized essentially by $\|q\|_{2/3}/\eps^2$. (More precisely, it is determined by the $2/3$-norm of 
$\overline{q}$ and $\underline{q}$, which are variants of $q$ formed by removing the largest and $\Theta(\epsilon)$ smallest probability masses from~$q$.) When $q$ is the uniform distribution, this recovers the worst-case $\Theta(\sqrt{d}/\epsilon^2)$ bounds for identity testing~\cite{paninski2008coincidence}.  

Similarly, our instance-optimal bounds are characterized by the Schatten-$\tfrac12$ quasinorm. To see this, note that
\begin{equation}
    F(\sigma, \mathbbm{1}/d) = \left\|\sqrt{\sigma} \sqrt{\mmstate}\right\|_1^2 
    = \frac{1}{d} \cdot \tr(\sqrt{\sigma})^2 = \frac{\|\sigma\|_{1/2}}{d}.
\end{equation}
Thus, the bounds in \Cref{thm:main-result} can be reformulated as
\begin{equation}
\label{eq:overview-quasinorm-result}
    \Tilde{\Omega} \left(\frac{\|\underline{\sigma}\|_{1/2}}{\epsilon^2}\right) \leq n \leq \Tilde{\bigo} \left(\frac{\|\overline{\sigma}\|_{1/2}}{\epsilon^2}\right),
\end{equation}
and this is the formulation we will consider throughout.

Ignoring both the $\eps$ dependency and other technical details, a natural overarching plan for analyzing instance-optimal certification would be the following:
\begin{enumerate}
    \item Show that certification for  $\mmstate$ requires $\Theta(d)$ copies (or $\Theta(\sqrt{d})$ samples, in the classical case). \label{step:1}
    \item Extend the lower bound to the more ``robust'' case when $\sigma$'s eigenvalues are within a constant factor of $\frac1d$; say, each is in $[\frac{.5}{d}, \frac{2}{d}]$.\label{step:2}
    \item Infer that when $\sigma$ has $r$ nonzero eigenvalues, all near $\frac{1}{r}$ (the ``rank-$r$ model case''), the copy complexity is $\Theta(r) = \Theta(\|\sigma\|_{1/2})$ (or in the classical case, $\Theta(\sqrt{r})$).\footnote{This explains the $\|\sigma\|_{1/2}$ appearing in \Cref{eq:overview-quasinorm-result}, as $p = 1/2$ is the $p$-norm such that $\|(\frac1r, \dots, \frac1r)\|_{p} = r$.  Similarly, in the classical case, $p = 2/3$ is the $p$-norm such that $\|(\frac1r, \dots, \frac1r)\|_{p} = \sqrt{r}$, explaining the $2/3$-norm arising in~\cite{valiant2017automatic}.} Generally, the lower bound won't require more than Step~2, but the upper bound may require more work.\label{step:3}
    \item \label{step:4}Group $\sigma$'s eigenvalues $\lambda_1, \dots, \lambda_d$ into  $\bigo(\log d)$ ``buckets'', each covering a constant multiplicative range.  The bucket with largest  contribution to $\|\sigma\|_{1/2}$ causes the lower bound (this contribution equals $\|\sigma\|_{1/2}$ up to log factors). For the upper bound, one hopes to test the $\bigo(\log d)$ buckets separately.
\end{enumerate}

The most important technical contribution of this work comes in achieving Step~\eqref{step:2} (note that \Cref{thm:intro-almost-mixedness}'s $\|\sigma^{-1}\|_2$ equals $\Theta(d^{3/2})$ in Step~\eqref{step:2}'s setting).  Having this improvement to Step~\eqref{step:1} is essential, as the subsequent bucketing step can only guarantee eigenvalues that are multiplicatively close to each other, not identical.

In the first part of this overview, \Cref{sec:techniques-prior}, we describe the only previously known proof of Step~\eqref{step:1}, and why it seems very difficult to make robust.  We also discuss the standard classical lower bound, which forms the basis for our quantum approach. Then, we provide an overview of our new techniques and how we use them to prove all our lower bounds in \Cref{sec:tech-lower-us}.  Finally, in \Cref{sec:tech-upper} we discuss Steps~\eqref{step:3} and~\eqref{step:4}, finding that the upper bound in \Cref{thm:main-result} follows from a straightforward modification of the unentangled-measurements algorithm of \cite{chen2022toward}.

\subsection{Prior Testing Lower Bounds}
\label{sec:techniques-prior}
Let us first recall the (tight) quantum lower bound of $\Omega(d/\eps^2)$ for mixedness testing with entangled measurements; i.e., certification for $\sigma = \mmstate = \diag(\frac1d, \dots, \frac1d)$. This lower bound, from~\cite{o2015quantum}, uses the representation-theory approach initiated by~\cite{childs2007weak}. A key first insight for this lower bound is that for any quantum state $\rho$, its distance $\|\rho -\mmstate\|_1$ from the maximally mixed state depends only on the spectrum of $\rho$. In other words, mixedness testing is a so-called \emph{spectrum testing} task. For such tasks, algorithms using \emph{weak Schur sampling} and purely classical post-processing are known to be optimal~\cite[Lemma~20]{montanaro2016survey}. This meant it sufficed for \cite{o2015quantum} to analyze a purely classical testing problem, namely distinguishing the ``Schur--Weyl distribution'' on $n$-box Young diagrams in the two parameter cases of $(\frac1d, \dots, \frac1d)$ and $(\frac{1+\eps}{d}, \frac{1-\eps}{d}, \dots, \frac{1+\eps}{d}, \frac{1-\eps}{d})$. This required some fairly sophisticated results, including the ``Binomial Formula for shifted Schur polynomials'' and expressions for the Schur polynomial and symmetric group characters at $(+1, -1, \dots, +1, -1)$.

Unfortunately, generalizing this result to the Step~\eqref{step:2} ``robust'' case is a complete non-starter, because when $\sigma \neq \mmstate$, state certification is no longer a spectrum testing task. 
Thus even to handle ``nearly maximally mixed'' states~$\sigma$, we need to develop a completely new methodology for proving lower bounds against testers that use entangled measurements.

\paragraph{Uniformity Testing via $\chi^2$.} 
Our new technique will be a quantum analogue of the widely successful Ingster(--Suslina) \cite{ingster2012nonparametric} method in classical distribution testing. In preparation for describing our methods, we first review how this technique can be used to get a tight lower bound for uniformity testing in the classical setting. 

Paninski~\cite{paninski2008coincidence} first showed that $n = \Omega(\sqrt{d}/\epsilon^2)$ samples are necessary to test identity to the uniform distribution $u$ over $\{1, \dots, d\}$. 
Paninski's proof starts by constructing a parameterized family of distributions $p_{\theta}$ such that $\dtv(p_{\theta}, u) \geq \epsilon$.  (Specifically, each $p_{\theta}$ is of the form $(\frac{1 \pm \eps}{d}, \frac{1 \mp \eps}{d}, \frac{1 \pm \eps}{d}, \frac{1 \mp \eps}{d}, \dots)$, where $\theta \in \{\pm 1\}^{d/2}$ specifies how the signs are chosen in each consecutive pair.)
In the proof, $\theta$ will be chosen uniformly at random.
Using a standard argument (``Le Cam's two-point method''), one can see that a successful tester must have $n$ large enough so that
\begin{equation}
    \dtv\left(
    \mbb{E}_{\bftheta} p_{\bftheta}^{\otimes n}, u^{\otimes n}
    \right) = \Omega(1).
\end{equation}
Thus it suffices to upper-bound the above total variation distance as a function of $n$. A crucial idea for achieving this is to pass to $\chi^2$-divergence, using the standard inequality $\dtv \leq \tfrac12 \sqrt{\divchi} $. Thus it suffices to lower-bound the $n$ required to achieve
\begin{equation}
    \divchi\left(
    \mbb{E}_{\bftheta} p_{\bftheta}^{\otimes n} \bigm\| u^{\otimes n}
    \right) = \Omega(1).
\end{equation}
 Although upper-bounding $\chi^2$-divergence is only harder than upper-bounding $\dtv$, it has the advantage that one can write exact formulas for $\divchi(\mbb{E}_{\bftheta} p_{\bftheta}^{\otimes n}\mathrel{\|}u^{\otimes n})$. This is the Ingster--Suslina method, which states
 \begin{equation}
     \divchi(\mbb{E}_{\bftheta} p_{\bftheta}^{\otimes n} \mathrel{\|}  u^{\otimes n}) = \mbb{E}_{\bftheta, \bftheta^\prime}\left[ (1 + H(\bftheta, \bftheta^\prime))^n\right] - 1,
 \end{equation}
where $\bftheta,\bftheta^\prime$ are i.i.d, and
 \begin{equation}
        H(\theta, \theta^\prime) \triangleq \mbb{E}_{\bx \sim u} \left[
            \frac{(p_{\theta}(\bx) - u(\bx))(p_{\theta^\prime}(\bx) - u(\bx))}{u(\bx)^2}
        \right].
 \end{equation} 
 It is not hard to rewrite the right-hand side above in terms of moment generating functions, and then bound it using large deviation results (e.g., Hoeffding's bound).  Precisely, one can show
 \begin{equation}
     \divchi(\mbb{E}_{\bftheta} p_{\bftheta}^{\otimes n} \mathrel{\|} u^{\otimes n}) = \bigo(n^2\epsilon^4/d),
 \end{equation}
 which results in the $\Omega(\sqrt{d}/\epsilon^2)$ bound. For detailed proofs along these lines, we refer the reader to Theorem~3.2 of \cite{canonne2022topics} and Section~24.3 of \cite{wu2017lecture}.
 
 We close by remarking that (although we haven't seen it in print) this proof \emph{is} reasonably easy to ``robustify''.  That is, one can make it work for distributions $u$ with all probabilities in, say, $[\frac{.5}{d}, \frac{2}{d}]$, by using suitable perturbations $p_\theta$ and relying again on Hoeffding's inequality.
 
\subsection{Our Lower Bounds}
\label{sec:tech-lower-us}
Our main lower bound technique is a direct quantum analogue of the classical arguments discussed above for uniformity testing. To prove lower bounds for certifying a state $\sigma$, we construct suitable parameterized states $\rho_{\bftheta}$ for a random parameter $\bftheta$, such that the $\rho_{\bftheta}$'s are all $\eps$-far from $\sigma$. Then we employ the crucial step of relating trace distance to $\qdivchi$, the \emph{quantum} $\chi^2$-divergence.  (In fact, to one of the many natural quantum analogues of $\chi^2$-divergence; see \Cref{def:quantum-chi2}.) As before, it follows that the copy complexity $n$ must be large enough to satisfy
\begin{equation}
    \qdivchi\left(\mathbb{E}_{\bftheta} \rho_{\bftheta}^{\otimes n} \bigm \| \sigma^{\otimes n}\right) \geq \Omega(1).
\end{equation}
 Analogously to the classical Ingster--Suslina method, we are able to explicitly compute (and then bound) this divergence as:
 \begin{equation}
     \qdivchi\left(\mathbb{E}_{\bftheta} \rho_{\bftheta}^{\otimes n} \bigm \| \sigma^{\otimes n}\right) = \mbb{E}_{\bftheta, \bftheta^\prime} \left[
        (1 + Z(\bftheta,\bftheta^\prime)^n)
    \right] -1 \leq \mbb{E}_{\bftheta, \bftheta^\prime} \left[\exp(n \cdot Z(\bftheta,\bftheta^\prime))\right] -1,
 \end{equation}
 where 
 \begin{equation}
     Z(\bftheta,\bftheta^\prime) = \tr(\sigma^{-1} (\rho_{\bftheta} - \sigma)(\rho_{\bftheta^\prime} - \sigma)).
 \end{equation}
It remains to bound the moment generating function-type expression above, which we do using concentration of measure techniques for random unitary matrices.  These work well as long as our alternative states $\sigma_{\bftheta}$ have bounded eigenvalues. In particular, we do not particularly need to have $\sigma = \mmstate$ precisely (though this does slightly simplify the argument); i.e., we are able to complete ``Step~\eqref{step:2}'', getting a lower bound for certifying nearly maximally mixed states.

Overall, our ``quantum Ingster--Suslina'' method proves to be straightforward to execute, yet surprisingly powerful, allowing us to prove many of our lower bounds with a few easy steps.

\paragraph{Mixedness Testing via Quantum $\chi^2$.} 
Here we give a few more details on how to execute the above proof when $\sigma = \mmstate$.  (As mentioned, this special case is mildly simpler than the case of nearly-maximally mixed states.)
We use the ``quantum Paninski construction'', which has previously been used in various forms for this task \cite{o2015quantum,bubeck2020entanglement,chen2021hierarchy}. Let $\Sigma = \diag(+1, -1, \dots, + 1, -1)$ be a diagonal $d \times d$ matrix with trace 0, with the last entry zeroed out if $d$ is odd. Then we consider the  parameterized family of states
\begin{equation}
    \rho_{\bfu} \triangleq \mmstate + \frac{\epsilon}{d}\bfu \Sigma \bfu^\dag,
\end{equation}
with $\bfu$ drawn from the Haar measure over the group of unitary $d \times d$ matrices. 
Then our quantum Ingster--Suslina method requires us to bound
\begin{equation}    \label{eqn:Z}
    \mathbb{E}_{\bfu,\bfv} [\exp(n \cdot Z(\bfu,\bfv))] -1, \quad \text{where} \quad 
    Z(\bfu,\bfv) = \tr( \left(\frac{\mathbbm{1}}{d}\right)^{-1} \left(\frac{\eps}{d} \bfu \Sigma \bfu^\dag\right) \left(\frac{\eps}{d} \bfv \Sigma \bfv^\dag \right)).
\end{equation}
Some simple manipulations show that this is equivalent to bounding the moment generating function of a mean-zero, degree-$2$ polynomial in the entries of a Haar-random unitary.  
Then standard concentration inequalities imply
\begin{equation}
   \qdivchi\left(
        \mathbb{E}_{\bfu}\rho_{\bfu}^{\otimes n} \bigm \| \left(\mmstate\right)^{\otimes n}
    \right) = \bigo\left(\frac{n^2\epsilon^4}{d^2}\right),
\end{equation}
which immediately recovers the $\Omega(d/\epsilon^2)$ lower bound for mixedness testing and proves \Cref{thm:intro-mixedness}.

\paragraph{Extending to Nearly Mixed States of Rank~$r$.} As discussed earlier, we now wish to extend the above ideas to states that are nearly maximally mixed. As mentioned, this is hardly much more difficult.  We can use the same perturbation family $\rho_{\bfu} \triangleq \sigma + \frac{\epsilon}{d}\bfu \Sigma \bfu^\dag$, and the method leads us to a mildly more complicated version of $Z(\bfu,\bfv)$ from \Cref{eqn:Z}, in which $({\mmstate})^{-1}$ is replaced by~$\sigma^{-1}$.  This does not make bounding the moment generating function much more difficult, and in this way we obtain \Cref{thm:intro-almost-mixedness}. The same  methodology also works well for states $\sigma$ that are nearly mixed of rank $r < d$ (the ``rank~$r$ model case'' discussed in Step~\eqref{step:3}). One constructs the mixture of alternatives $\rho_{\bfu}$ by perturbing only $\sigma$'s dimension-$r$ support subspace by $\frac{\eps}{r} \bfu \Sigma \bfu^\dagger$.

\paragraph{Handling General $\sigma$.} While the above methods work well for nearly mixed states of any rank, we cannot use them to get instance-optimal bounds in general. First, if $\sigma$ has some small nonzero eigenvalue $\lambda_{\min}$, this limits the range of $\epsilon$ for which our bound can be applied, as $\lambda_{\min}$ must be at least $\epsilon/r$ to ensure that $\rho_{\bfu}$ is positive semidefinite. Suppose we were to settle for a limited range of $\eps$ and disregard this first concern. Even then, the bound would not be optimal, as this construction places too much emphasis on the small eigenvalues of~$\sigma$. For instance, let $\sigma$ have $d-1$ eigenvalues that are $\Omega(1/d)$, and let the last eigenvalue be~$1/d^2$. For certifying such a state, one would still expect an $\Omega(d/\eps^2)$ lower bound. However, $\|\sigma^{-1}\|_2^2 = $ \mbox{$\bigo((d-1) \cdot d^2 + d^4) = \bigo(d^4)$}, where we see that the contribution of the smallest eigenvalue drowns out the rest. Then, substituting into \Cref{thm:intro-almost-mixedness} only yields a lower bound of $\Omega(\sqrt{d}/\eps^2)$ instead of the desired $\Omega(d/\eps^2)$ bound. 

While one strategy to overcome this issue would be to entirely neglect the small eigenvalues and only perturb the subspace corresponding to large ones, this runs the risk of neglecting too many eigenvalues, which would also yield sub-optimal bounds. 
Instead, the strategy that works is to \emph{bucket} the eigenvalues of~$\sigma$ into logarithmically many ranges with similar magnitude, and separately perturb the eigenspaces corresponding to each bucket.

\paragraph{Bucketing and Block-Diagonal Perturbations.} 
We now give an overview (omitting some minor details) of how bucketing is used to obtain the instance-optimal lower bound for general~$\sigma$. 
We first neglect suitably chosen eigenvalues of~$\sigma$ adding up to $\bigo(\epsilon)$, and then group the remaining eigenvalues into buckets, with bucket $S_j$ containing all eigenvalues between $2^{-j-1}$ and $2^{-j}$. Let the set of buckets be $\mc{J}$ and the number of buckets be $m = |\mc{J}|$. Let $\sigma_j$ be the principal submatrix of $\sigma$ associated with $S_j$, and let $d_j = |S_j|$. Further, let $\sigma_{\mrm{tail}}$ be the principal submatrix associated with the neglected eigenvalues. Then, we individually perturb each such submatrix to define
\begin{equation}
    \sigma_{(\bfu_1, \dots, \bfu_m)} \triangleq \sigma_{\mrm{tail}} \oplus \bigoplus_{j \in \mc{J}} (\sigma_j + \epsilon_j \bfu \Sigma_j \bfu^{\dagger}),
\end{equation}
where $\bfu_j$ are independently drawn from the Haar measure over the group of $d_j \times d_j$ unitary matrices, $\Sigma_j$ is a $d_j \times d_j$ matrix defined similarly to $\Sigma$, and the choice of $\epsilon_j$'s will be explained later. Using our quantum Ingster--Suslina method, we are led to analyze
\begin{equation}
    Z(
    \bfu_1, \dots, \bfu_m, \bfv_1, \dots, \bfv_m
    ) = \tr(\sigma^{-1} (\sigma_{(\bfu_1, \dots, \bfu_m)} - \sigma)(\sigma_{(\bfv_1, \dots, \bfv_m)} - \sigma)).
\end{equation}
  The block diagonal structure of our constructed states $\sigma_{(\bfu_1, \dots, \bfu_m)}$ allows us to conveniently rewrite the above quantity as a sum of independent random variables $Z_j (\bfu_j, \bfv_j)$ associated with each bucket, with
\begin{equation}
     Z_j (\bfu_j, \bfv_j) \triangleq \epsilon_j^2 \tr(\sigma_j^{-1} \bfu_j \Sigma_j \bfu_j^\dagger \bfv_j \Sigma_j \bfv_j^\dag).
\end{equation}
Then, by the independence of these random variables, we get
\begin{equation}
    \mathbb{E} \exp\Bigl( n \sum_j Z_j (\bfu_j, \bfv_j)\Bigr) = \prod_j \mathbb{E}_{\bfu_j, \bfv_j} \exp(n Z_j (\bfu_j, \bfv_j)).
\end{equation}
Having decoupled the analysis across buckets, we are able to bound each of these individual expectations above using Haar measure concentration arguments, similarly to the near-mixedness testing case. This yields the following lower bound:
\begin{equation}
    n = \Omega\left(\sum_j \eps_j^4 2^{2j}\right)^{-1/2}.
\end{equation}
Similar-looking perturbation-dependent lower bounds appear in prior work on instance-optimal identity testing \cite[Theorem 4]{valiant2017automatic} and state certification from unentangled measurements \cite[Lemma 5.7]{chen2022toward}. By carefully tuning the perturbations $\eps_j$ using arguments similar to \cite{valiant2017automatic,chen2022toward}, we obtain the lower bound of \Cref{thm:main-result} except in one corner case, which we discuss below.

\paragraph{Corner Case: Large Principal Eigenvalue.}
When the above lower bound does not yield that in \Cref{eq:overview-quasinorm-result}, \cite{chen2022toward} showed that two important conditions hold. First, the largest eigenvalue of $\sigma$ is at least $\tfrac12$. Second, $\|\underline{\sigma}\|_{1/2} = \Tilde{\bigo}(1)$. Then, to obtain our desired lower bound, it suffices to show an $\Omega(1/\epsilon^2)$ lower bound for certifying states with large principal eigenvalues. We split this proof across two cases, depending on whether or not the second-largest eigenvalue is below~$\tfrac14$. 

In the former case, when the second eigenvalue is at most $\tfrac14$, we take inspiration from the instance used to prove the lower bound for testing \emph{equality} of two pure states by \cite{montanaro2016survey}. Given $\sigma$, we construct an alternative state $\sigma^\prime$ by rotating the eigenvectors of $\sigma$ associated with the two largest eigenvalues. We then obtain the desired $\Omega(1/\epsilon^2)$ bound by upper bounding the fidelity $F(\sigma^{\otimes n},\sigma^{\prime\otimes n})$, and relating fidelity to the trace distance.

On the other hand, when the second eigenvalue is at least $\tfrac14$, we consider an instance from \cite{chen2022toward}. Here, the alternative states are obtained by randomly perturbing the $2 \times 2$ principal submatrix of $\sigma$ associated with these two largest eigenvalues. We then obtain the desired lower bound by applying the quantum Ingster--Suslina method to this instance. 

\subsection{Upper Bound}
\label{sec:tech-upper}
We obtain our upper bound in \Cref{eq:overview-quasinorm-result} by simply replacing a key subroutine of the unentangled instance-optimal tester from \cite{chen2022toward} with an entangled one. The main idea behind their algorithm is to use bucketing to break down the task into simpler testing tasks, and then use an unentangled Hilbert--Schmidt tester to solve each of them. 

We now sketch the method in some more detail. Once again, we group the eigenvalues of~$\sigma$ into buckets. First, we collect the bottom $\bigo(\epsilon^2)$ mass into a tail bucket $\stail$. Then, we bucket the remaining eigenvalues such that the bucket $S_j$ contains all eigenvalues between $2^{-j-1}$ and $2^{-j}$. Let the number of such buckets be $m$, which is at most $\bigo(\mathrm{log}(d/\eps))$ (see \Cref{def:bucketing-upper}). Instead of \cite{chen2022toward}'s unentangled Hilbert--Schmidt certification algorithm, we will use the algorithm from \cite[Theorem 1.4]{buadescu2019quantum}, which we refer to as \textsf{HSCertify}. Specifically, \textsf{HSCertify} can distinguish between $\rho = \sigma$ and $\|\rho - \sigma\|_{\HS} \geq \epsilon$ with high probability using $\bigo(1/\epsilon^2)$ copies. Now \cite{chen2022toward} essentially showed that if $\dtr(\rho,\sigma) > \epsilon$, then one of three cases must occur. We describe the cases and how we handle them as follows: 

\begin{enumerate}
    \item \textbf{Heavy Tail}: The easiest case to test for is when the principal submatrix of $\rho$ corresponding to the indices in $\stail$ has $\Omega(\epsilon^2)$ more mass than that of $\sigma$. In this case, we perform projective measurements on individual copies of $\rho$, allowing us to handle the case with $\bigo(1/\epsilon^2)$ measurements.
    \item \textbf{Block-Diagonal Perturbations:} In this case, there exists some bucket $S_j$ for which the corresponding principal submatrices of $\rho$ and $\sigma$ are $\Omega(\epsilon/m^2)$-far. Then, we test for such an error across each bucket by performing some simple preprocessing, projecting into the bucket, and then passing to \textsf{HSCertify}. This allows us to test for this case with $\Tilde{\bigo}(\|\overline{\sigma}\|_{1/2}/\epsilon^2)$ copies of $\rho$.
    \item \textbf{Off-Diagonal Perturbations:} In this case, there exists a pair of buckets for which the corresponding \emph{off-diagonal} submatrices of $\rho$ and $\sigma$ are $\Omega(\epsilon/m^2)$-far. We test for such errors across each pair of buckets, again by performing some preprocessing, then projecting, and finally passing to \textsf{HSCertify}. Once again, this allows us to test for this case with $\Tilde{\bigo}(\|\overline{\sigma}\|_{1/2}/\epsilon^2)$ copies of~$\rho$. 
\end{enumerate}

By testing for the above cases, we can $\eps$-certify~$\sigma$ with the nearly-optimal copy complexity~$\Tilde{\bigo}(\|\overline{\sigma}\|_{1/2}/\epsilon^2)$.

\paragraph{Why We Don't Need an Off-Diagonal Instance.}
We highlight that our tests for Cases 2 and 3 above have the same copy complexity. In the setting of unentangled measurements, the certification algorithm of \cite{chen2022toward} also tests for the same three cases. However, in their setting, the overall copy complexity is solely dominated by the tester for Case~3. In other words, with unentangled measurements, testing for \emph{off-diagonal} perturbations is harder than testing for \emph{diagonal} ones. Our results demonstrate a contrast in the entangled setting, where our tests for diagonal and off-diagonal perturbations have the same copy complexity.

This comparison is also reflected in the lower bounds. \cite{chen2022toward,chen2022tightStateCertification} mainly obtained their optimal lower bounds using an instance constructed from \emph{off-diagonal} perturbations, except for certain parameter regimes. However with entangled measurements, we show that an instance constructed from block-diagonal perturbations yields optimal lower bounds throughout. This reaffirms that with entangled measurements, the copy complexity of testing for block-diagonal perturbations turns out to be as much as that for full state certification. As analyzing the diagonal instance suffices, we do not consider lower bounds for testing for off-diagonal perturbations. This makes our lower-bound casework simpler than that of \cite{chen2022toward,chen2022tightStateCertification}, and also allows our mass removal to be somewhat lighter. 

\subsection*{Organization}
In \Cref{sec:prelim}, we introduce basic notation and necessary technical background. In \Cref{sec:quantum-ingster}, we present our quantum Ingster--Suslina method and use it to prove some preliminary lower bounds, including \Cref{thm:intro-mixedness,thm:intro-almost-mixedness}. We prove our main lower bound from \Cref{thm:main-result} in \Cref{sec:lower} and the upper bound in \Cref{sec:upper}.

\subsection*{Acknowledgments} 
The authors thank Francesco Anna Mele, Jinge Bao, Sitan Chen, Mina Doosti, Weiyuan Gong, and Laura Lewis for valuable discussions at various stages of this project. RO was supported    in part by a grant from Google Quantum AI.
CW was supported by the UK EPSRC through the Quantum Advantage Pathfinder project with grant reference EP/X026167/1. 
\section{Preliminaries}
\label{sec:prelim}
For an integer $n \geq 1$, let $[n]$ denote $\{1, 2, \dots, n\}$. Unless explicitly stated otherwise, we will let $d$ denote the dimension of the quantum states under consideration. A quantum state $\rho$ can be associated with a density matrix, which is a Hermitian positive semidefinite operator on $\mbb{C}^{d}$ with unit trace. Let
$\mmstate$ denote the maximally mixed state, where $\mathbbm{1}$ is the $d \times d$ identity operator. For $p > 0$, let $\|\cdot\|_p$ denote the Schatten-$p$ norm, which is a quasi-norm for $p < 1$. Let $U(d)$ denote the group of unitary $d \times d$ matrices. Throughout, we will use $\Tilde{\Omega}, \Tilde{\bigo},$ and  $\Tilde{\Theta}$ to suppress $\mathrm{polylog}(d/\eps)$ factors.

\subsection{Distances and Divergences}
We will define distances and divergences between probability distributions as well as quantum states; and, we state some of their useful properties. Given two distributions $p,q$ over a discrete domain $\mc{X}$, their total variation distance is $\dtv(p,q) \triangleq \frac12 \|p-q\|_1$. The $\chi^2$-divergence between the distributions is given by
\begin{equation}
    \divchi(p\|q) \triangleq \sum_{x \in \mc{X}} \frac{(p(x) - q(x))^2}{q(x)},
\end{equation}
which is infinite if $p$ is supported on any element not in the support of $q$. The total variation distance and $\chi^2$-divergence are related by the following inequality:
\begin{equation}
\label{eq:dtv-divchi}
    \dtv(p,q) \leq \tfrac12\sqrt{ \divchi(p\|q)}.
\end{equation}

We will now define important distances between quantum states. Analogous to the total variation distance, the trace distance between quantum states $\rho,\sigma \in \mbb{C}^{d \times d}$ is defined to be $\dtr(\rho,\sigma) \triangleq \frac12\|\rho - \sigma\|_1$, where recall $\|\cdot\|_1$ is the Schatten 1-norm. 

We also define a quantum analogue of the $\chi^2$-divergence, originally presented in \cite{temme2010chi}. While \cite{temme2010chi} presented a family of quantum $\chi^2$-divergences, we will find it convenient to restrict our attention to the largest such divergence, defined as follows:
\begin{definition}[Quantum $\chi^2$-divergence, \cite{temme2010chi}]
\label{def:quantum-chi2}
Let $\rho, \sigma \in \mbb{C}^{d \times d}$ be quantum states. 
In case $\sigma$ is nonsingular, we define the quantum $\chi^2$-divergence of $\rho$ from $\sigma$ as 
    \begin{equation}
        \qdivchi(\rho||\sigma) = \tr(\sigma^{-1}(\rho-\sigma)^2) = \tr(\sigma^{-1}\rho^2) - 1.
    \end{equation}
For singular~$\sigma$, the $\chi^2$-divergence is taken to be~$\infty$ if $\ker \sigma \not \leq \ker \rho$; otherwise, we use the above definition but restrict $\sigma, \rho$ to $\mathop{\textnormal{Im}} \sigma$.
\end{definition}
Our lower bounds will make use of the following inequality relating the trace distance and the quantum $\chi^2$-divergence.
\begin{lemma}[Lemma 5 in \cite{temme2010chi}]
\label{lem:dtr-qdivchi}
    Let $\rho, \sigma \in \mbb{C}^{d \times d}$ be quantum states. Then
    \begin{equation}
        \dtr(\rho,\sigma) \leq \tfrac12 \sqrt{\qdivchi(\rho||\sigma)}.
    \end{equation}
\end{lemma}

It will also be useful to define the fidelity between two quantum states, $F(\rho, \sigma) \triangleq 
\|\sqrt{\rho}\sqrt{\sigma}\|_1^2$, a measure of their closeness. 
We will use the following fidelity formula, which holds for qubits:

\begin{lemma}[Equation (10) in \cite{jozsa1994fidelity}]
\label{lem:fid-jozsa}
    Given quantum states $\rho, \sigma \in \mathbb{C}^{2 \times 2}$ with Bloch-vector representations $\vec{a}$ and $\vec{b}$ respectively, we have
    \begin{equation}
        F(\rho, \sigma) = \frac12\left(1+\vec{a} \cdot \vec{b} +\sqrt{(1-\vec{a}\cdot \vec{a})(1-\vec{b}\cdot \vec{b})}\right).
    \end{equation}
\end{lemma}
We will also make use of the following properties of the fidelity:
\begin{lemma}[Section 9.2.3 of~\cite{nielsen2010quantum}]
\label{lem:dtr-infidelity}
    Given quantum states $\rho, \sigma \in \mathbb{C}^{d \times d}$, we have 
    \begin{equation}
        \dtr(\rho, \sigma) \leq \sqrt{1 - F(\rho,\sigma)}.
    \end{equation}
\end{lemma}

\begin{lemma}[Tensorization of Fidelity, Section 5.P5 of \cite{jozsa1994fidelity}]
    \label{lem:fidelity-tensorization}
    Given quantum states $\rho, \sigma \in \mathbb{C}^{d \times d}$, we have 
    \begin{equation}
        F(\rho^{\otimes n}, \sigma^{\otimes n}) = F(\rho, \sigma)^n.
    \end{equation}
\end{lemma}

\subsection{Unitary Haar Measure}
In this section, we present some useful properties of the Haar measure over the unitary group $U(d)$. 
We first state an elementary and well-known fact:

\begin{fact}
    \label{fact:haar-measure-moments}
    Let $\Sigma \in \mbb{C}^{d \times d}$. 
    Then $\mbb{E}_{\bfu \sim U(d)}[\bfu \Sigma \bfu^\dag] = \tr(\Sigma) \cdot \mmstate$.
\end{fact}
\begin{proof}
    This follows directly from $\mbb{E}[\bfu_{ij} \overline{\bfu}_{ik}]$ being $1/d$ if $j = k$, and $0$ otherwise.
\end{proof}

Next, we state a useful concentration-of-measure inequality for the Haar measure.

\begin{definition}[Lipschitzness.]
    Let $f : U(d)^k \to \mbb{R}$.  
    We will say $F$ is $L$-Lipschitz (with respect to the $L_2$-sum of Hilbert--Schmidt metrics) if $|f(U_1, \dots, U_k) - f(V_1, \dots, V_k)| \leq L \cdot \sqrt{\sum_{i=1}^k \|U_i - V_i\|_2^2}$.
\end{definition}

\begin{lemma}
\label{lem:mgf-upper}
    Let $f : U(d)^k \rightarrow \mbb{R}$ 
    have mean zero; i.e., assume $\mbb{E}_{\bfu_1, \dots, \bfu_k}[f(\bfu_1, \dots, \bfu_k)] = 0$ for $\bfu_1, \dots, \bfu_k$ drawn independently from the Haar measure on $U(d)$.
    Assume also that $f$ is $L$-Lipschitz. 
    Then
     \begin{equation}
        \mbb{E}[\exp(f(\bfu_1, \dots , \bfu_k))] \leq \exp(3L^2/d).
    \end{equation}
\end{lemma}
\begin{proof}
    As shown by Meckes and Meckes~\cite[Corollary~17]{meckes2013spectral}, $f$'s domain satisfies a log-Sobolev inequality with constant $C = 6/d$.
    The result now follows from the so-called ``Herbst argument''; precisely, one may take $F = f/L$, and $\lambda = L$ in \cite[inequality (2.19)]{ledoux1999concentration}.

\end{proof}
\section{A Quantum Ingster--Suslina Method}
\label{sec:quantum-ingster}
In this section, we establish our quantum Ingster--Suslina method, as described in \Cref{sec:techniques-prior,sec:tech-lower-us}.

We start by recalling the following fact concerning distinguishing two mixed states, which is sometimes called Helstrom's Lemma:
\begin{fact}\label{fact:quantum-point-mixture-dtr} 
    Let $\sigma \in \mbb{C}^{d \times d}$ be a quantum state and let $(\rho_{\theta})_{\theta}$ be a family of such states, parameterized by a random variable~$\bftheta$. Given $n$ copies of an unknown state $\rho$, where either $\rho = \sigma$ or $\rho = \rho_{\bftheta}$, any algorithm can distinguish between the two cases with probability at most $\frac12 + \frac12 \dtr\left(
             \mbb{E}_{\bftheta} \rho_{\bftheta}^{\otimes n}, \sigma^{\otimes n}
        \right).$
\end{fact}
Next, using \Cref{lem:dtr-qdivchi}, we could replace the right-hand side above with $\frac12 + \frac14\sqrt{\qdivchi(\mbb{E}_{\bftheta} \rho_{\bftheta}^{\otimes n} \|\sigma^{\otimes n})}$.  In particular, if the $\chi^2$-divergence is at most $4/9$, say, then this quantity is at most~$2/3$. We conclude:
\begin{fact} \label{fact:quantum-point-mixture-chi} In the setting of \Cref{fact:quantum-point-mixture-dtr}, 
if there exists an algorithm distinguishing between $\rho = \sigma$ and $\rho = \rho_{\bftheta}$ with high probability, then
\begin{equation}
    \qdivchi(\mbb{E}_{\bftheta}\rho_{\bftheta}^{\otimes n} \| \sigma^{\otimes n}) \geq  \Omega(1).
\end{equation}   
\end{fact}
We can now establish the following:

\begin{lemma}[A quantum Ingster--Suslina method]
\label{lem:quantum-ingster-suslina}
    Let $\sigma \in \mathbb{C}^{d \times d}$ be a quantum state, and let $(\rho_\theta)_{\theta}$ be a family of such states, parameterized by a random variable $\bftheta$. Then
    \begin{align}
        \qdivchi(\mbb{E}_{\bftheta} \rho_{\bftheta}^{\otimes n} \| \sigma^{\otimes n}) = \mbb{E}_{\bftheta, \bftheta^\prime}\tr(\sigma^{-1}\rho_{\bftheta} \rho_{\bftheta^\prime})^n - 1 
    &= \mbb{E}_{\bftheta, \bftheta^\prime} \left(1 + \tr(\sigma^{-1}(\rho_{\bftheta} - \sigma)(\rho_{\bftheta^\prime} - \sigma))\right)^n - 1 \\
    &\leq \mbb{E}_{\bftheta, \bftheta^\prime} \exp\left(n \cdot \tr(\sigma^{-1}(\rho_{\bftheta} - \sigma)(\rho_{\bftheta^\prime} - \sigma))\right) - 1.\label{eqn:key}
    \end{align}
\end{lemma}
\begin{proof}
Using \Cref{def:quantum-chi2}, we have
\begin{align}
    \qdivchi(\mbb{E}_{\bftheta} \rho_{\bftheta}^{\otimes n} \| \sigma^{\otimes n}) &= \tr( (\sigma^{-1})^{\otimes n} (\mbb{E}_{\bftheta} \rho_{\bftheta}^{\otimes n})^2) - 1 
    \\&= \tr( (\sigma^{-1})^{\otimes n} (\mbb{E}_{\bftheta} \rho_{\bftheta}^{\otimes n}) (\mbb{E}_{\bftheta^\prime} \rho_{\bftheta^\prime}^{\otimes n})) - 1
    \\&= \mbb{E}_{\bftheta,\bftheta^\prime} \tr((\sigma^{-1})^{\otimes n} \rho_{\bftheta}^{\otimes n} \rho_{\bftheta^\prime}^{\otimes n})  - 1
    \\&= \label{eq:quantum-ingster-1} \mbb{E}_{\bftheta,\bftheta^\prime} \tr(\sigma^{-1} \rho_{\bftheta} \rho_{\bftheta^\prime})^n - 1,
\end{align}
thus proving the first equality in the lemma. The second equality follows from
\begin{align}
    \tr(\sigma^{-1} (\rho_{\bftheta}-\sigma) (\rho_{\bftheta^\prime}-\sigma)) &= \tr(\sigma^{-1} \rho_{\bftheta} \rho_{\bftheta^\prime}) - \tr(\sigma^{-1} \rho_{\bftheta} \sigma) - \tr(\sigma^{-1} \sigma \rho_{\bftheta^\prime})+\tr(\sigma^{-1} \sigma \sigma)
    \\&= \tr(\sigma^{-1} \rho_{\bftheta} \rho_{\bftheta^\prime}) - 1. 
\end{align}
The final inequality is $1+x \leq \exp(x)$.
\end{proof}
In the remainder of this section, we apply this quantum Ingster--Suslina method to prove some preliminary lower bounds. These proofs demonstrate the strength and simplicity of our method for proving quantum testing lower bounds, and they serve as a warmup for our main lower bound in \Cref{sec:lower}. First, in \Cref{sec:warm-up-mixedness}, we present a new proof of the lower bound for certifying $\mmstate$. For this problem, we recover the $\Omega(d/\epsilon^2)$ lower bound of \cite{o2015quantum} with a vastly simpler proof. Next, we extend this idea to lower bounds for certifying nearly maximally mixed states in \Cref{sec:warm-up-2}. Finally, we prove a lower bound based on bucketing in \Cref{sec:warm-up-3}. We will use this last result to prove our main lower bound in \Cref{sec:lower}.

\subsection{Warmup: Mixedness Testing}
\label{sec:warm-up-mixedness}
In this section, we prove a lower bound on Mixedness Testing. 
Throughout this section, fix $0 \leq \eps \leq 1$, fix an even\footnote{This is merely for notational convenience. An extremely minor modification handles odd~$d$; or, one can use \Cref{thm:almost-mixedness-testing2} from \Cref{sec:warm-up-2}.} dimension~$d$, and let $\Sigma  \triangleq \diag(-1,+1,\dots,-1,+1) \in \mathbb{C}^{d \times d}$.
Given a unitary $U \in \mbb{C}^{d \times d}$, define
\begin{equation}
\label{eq:quantum-paninski-instance0}
    \rho_U \triangleq \mmstate + \frac{\epsilon}{d} U \Sigma U^\dag,
\end{equation}
which is a density matrix satisfying 
\begin{equation}
    \left\|\mmstate - \rho_U\right\|_1 = \epsilon.
\end{equation}
We can thus show a lower bound on Mixedness Testing (distinguishing $\rho = \mmstate$ from $\|\rho - \mmstate\|_1 \geq \eps$) as follows:
\begin{theorem}[Mixedness Testing lower bound with $\qdivchi$]
    \label{thm:mixedness-lower0}
    Any algorithm that $\eps$-certifies~$\mmstate$ must use $\Omega(d/\epsilon^2)$ copies of the unknown state.
\end{theorem}
\begin{proof}
    Per \Cref{fact:quantum-point-mixture-chi}, it suffices to establish the following:
    \begin{equation}
        \qdivchi \Bigl(\mbb{E}_{\bfu} \rho_{\bfu}^{\otimes n} \bigm\| \left(\mathbbm{1}/d\right)^{\otimes n}\Bigr) \leq \exp\left(\frac{12 n^2 \eps^4}{d^2}\right) - 1,
    \end{equation}
    as the quantity on the right is at most~$4/9$ when $n \leq .1 d/\eps^2$.
    Using \Cref{lem:quantum-ingster-suslina}, we bound:
    \begin{align}
        \qdivchi \Bigl(\mbb{E}_{\bfu} \rho_{\bfu}^{\otimes n} \bigm\| \left(\mathbbm{1}/d\right)^{\otimes n}\Bigr) + 1
        &\leq \mathop{\mbb{E}}_{\bfu,\bfv} \exp\left(
        n \cdot \tr((\mathbbm{1}/d)^{-1} (\rho_{\bfu} - \mathbbm{1}/d)(\rho_{\bfv} - \mathbbm{1}/d))
        \right)
        \\
        &= \mathop{\mbb{E}}_{\bfu,\bfv} 
        \exp\left(\frac{n \eps^2}{d} \cdot 
        \tr(\bfu\Sigma \bfu^\dag \bfv \Sigma \bfv^\dag )\right) = 
        \mathop{\mbb{E}}_{\bfu} 
        \exp\left(\frac{n \eps^2}{d} \cdot 
        \tr(\bfu^\dag \Sigma \bfu \Sigma)\right);
    \label{eq:mixedness-lower-divergence-bound0}
    \end{align}
    here we used $\tr(\bfu\Sigma \bfu^\dag \bfv \Sigma \bfv^\dag) = \tr((\bfu^\dag \bfv)^\dag\Sigma (\bfu^\dag \bfv) \Sigma)$, and that $\bfu^\dag \bfv$ has the same distribution as~$\bfu$.
    Now in light of \Cref{lem:mgf-upper}, it suffices to show that $f(\bfu) \triangleq \tr(\bfu^\dag \Sigma \bfu \Sigma)$ has mean zero and is $2\sqrt{d}$-Lipschitz.
    For the mean zero condition, we have
    \begin{equation}
        \mbb{E}[f(\bfu)] 
        = \mbb{E}\tr(\bfu^\dag \Sigma \bfu \Sigma) = \tr\left(\mbb{E}[\bfu^\dag \Sigma \bfu] \Sigma\right) 
        = 
        \tr\left((\tr \Sigma)\cdot \mmstate \cdot
        \Sigma\right) = 0,
    \end{equation}
    where we used \Cref{fact:haar-measure-moments} and $\tr(\Sigma) = 0$.
    As for the $2\sqrt{d}$-Lipschitz condition, for $U,V \in U(d)$ we have
    \begin{align}
        |f(U) - f(V)| &= \left|\tr(U^\dagger \Sigma U \Sigma) - \tr(V^\dagger \Sigma V \Sigma) \right|
        \\
        &\leq \left|\tr(U^\dagger \Sigma U \Sigma) - \tr(V^\dagger \Sigma U \Sigma)\right| + \left|\tr(V^\dagger \Sigma U \Sigma) - 
        \tr(V^\dagger \Sigma V \Sigma) \right|
        \\&\leq \|U^\dag-V^\dag\|_{\HS} \|\Sigma U \Sigma\|_{\HS} + \|U-V\|_{\HS} \|\Sigma V^\dag \Sigma\|_{\HS} \\
        &= 2\sqrt{d}\cdot \|U - V\|_{\HS},
    \end{align}
    where  the second line used the triangle inequality, the third line used matrix H\"older (and cyclicity of trace), 
    and the final line used that $\Sigma U \Sigma$ and $\Sigma V \Sigma$ are unitary (since $\Sigma$ is unitary) and hence have Frobenius norm equal to~$\sqrt{d}$.
\end{proof}

\subsection{Generalizing to Near-Maximally Mixed States} \label{sec:warm-up-2}
The original $\Omega(d/\eps^2)$ lower bound for Mixedness Testing~\cite{o2015quantum} used fairly involved representation-theoretic formulas that heavily relied on the hypothesis state~$\sigma$ being precisely $\mmstate$.
By contrast, our proof of \Cref{thm:mixedness-lower0} was only very mildly simplified by having $\sigma = \mmstate$. 
For example, in this section we show that the same lower bound can be achieved for ``nearly-maximally mixed states'', meaning ones where all eigenvalues are~$\Omega(1/d)$.

Let us encapsulate the tools used in the previous section.
Defining $\bfDelta$ to be the ``perturbation matrix'' $\rho_{\bftheta} - \sigma$,
the key step was bounding the following expression from \Cref{eqn:key}:
\begin{equation}
    \mbb{E} \exp\left(n \cdot \tr(\sigma^{-1}\bfDelta\bfDelta^\prime)\right)
\end{equation}
(where $\bfDelta^\prime$ denotes an independent copy of~$\bfDelta$).
To use the concentration-of-measure bound from \Cref{lem:mgf-upper} on this, we first need the natural condition that the perturbation $\bfDelta$ has mean zero; i.e., that $\mbb{E}[\rho_{\bftheta}] = \sigma$. 
Next, in the natural case that $\bfDelta$ has a unitarily-invariant distribution, we show that a good bound can be achieved provided $\bfDelta$ has small operator norm:
\begin{lemma}   \label{lem:exp-bound}
    Let $\sigma \in \mbb{C}^{d \times d}$ be nonsingular, and let $\bfDelta$ be a random Hermitian matrix with unitarily-invariant distribution; precisely,  
    $
        \bfDelta = \bfu \Sigma \bfu^\dag
    $ 
    for $\bfu \sim U(d)$ having Haar distribution and $\Sigma = \diag(\delta_1, \dots, \delta_d)$.
    Assume:
    \begin{itemize}
        \item $\mbb{E} \bfDelta = 0$; equivalently (\Cref{fact:haar-measure-moments}), $\tr(\Sigma) = \sum_j \delta_j = 0$.
        \item $\|\bfDelta\|_{\oper} \leq \delta$ always; equivalently, $|\delta_j| \leq \delta$ for all~$j$.
    \end{itemize}
    Then for $\bfDelta^\prime$ an independent copy of $\bfDelta$, and any real parameter~$n$,
    \begin{equation}
        \mbb{E} \exp\left(n \cdot \tr(\sigma^{-1}\bfDelta\bfDelta^\prime)\right) \leq \exp\left(\frac{48 n^2 \cdot \|\sigma^{-1}\|_{\HS}^2 \cdot \delta^4}{d}\right).
    \end{equation}
    In particular, this is at most $1.02$ whenever $\displaystyle
        n \leq .02 \cdot \frac{\sqrt{d}}{\|\sigma^{-1}\|_{\HS} \cdot \delta^2}.$
\end{lemma}
\begin{proof}
    Note that we can push scaling factors from $\sigma$ and $\Sigma$ into the parameter~$n$; thus we we may assume without loss of generality that
    $
        \|\sigma^{-1}\|_2 = \delta = 1
    $ 
    and show 
    \begin{equation}
        \mbb{E} \exp\left(n \cdot \tr(\sigma^{-1}\bfDelta\bfDelta^\prime)\right) \leq \exp(48 n^2/d).
    \end{equation}
    Let $f(U,V) = \tr(\sigma^{-1} \cdot U \Sigma U^\dag \cdot V \Sigma V^\dag)$, so our goal is to bound
    $
        \mbb{E}_{\bfu, \bfv \sim U(d)} [\exp(n f(\bfu,\bfv))].
    $ 
    In light of \Cref{lem:mgf-upper}, it suffices to show that $f(\bfu,\bfv)$ has mean zero and has Lipschitz constant bounded by~$4$.
    First we verify the mean zero condition:
    \begin{equation}
        \mbb{E} f(\bfu, \bfv) = \tr(\sigma^{-1} \cdot \mbb{E}[\bfu \Sigma \bfu^\dag] \cdot \mbb{E}[\bfu \Sigma \bfu^\dag]) = \tr(\sigma^{-1} \cdot\mbb{E}[\bfDelta] \cdot \mbb{E}[\bfDelta^\prime]) = \tr(0) = 0.
    \end{equation}
    As for the $4$-Lipschitz condition, we will now show that for $U,V,U' \in U(d)$ we have
    \begin{equation} \label{eqn:verifyme0}
        |f(U,V) - f(U',V)| \leq 2\|U - U'\|_2.
    \end{equation}
    A very similar proof (left to the reader) shows that $|f(U',V)- f(U',V')| \leq 2\|V - V'\|_2$ as well, from which we may conclude that $f$ is $4$-Lipschitz, as desired.
    Now to verify \Cref{eqn:verifyme0}, we have
    \begin{align}
        |f(U,V) - f(U',V)| &= \left|\tr(\sigma^{-1} (U \Sigma U^\dag) V \Sigma V^\dag) - \tr(\sigma^{-1} (U' \Sigma {U'}^\dag) V \Sigma {V}^\dag)\right|
        \\
        &\leq \left|\tr(\sigma^{-1}   (U - U') \Sigma {U}^\dag V \Sigma V^\dag)\right| + \left|\tr(\sigma^{-1} U' \Sigma (U^\dag - {U'}^\dag) V \Sigma {V}^\dag )\right|
        \\
        &\leq \|U-U'\|_{\HS} \cdot \|\Sigma U^\dag V \Sigma V^\dag \sigma^{-1}\|_{\HS} + \|U - U'\|_{\HS} \cdot \|V \Sigma V^\dag \sigma^{-1} U' \Sigma\|_{\HS} \\
        &\leq \|U-U'\|_{\HS}  + \|U-U'\|_{\HS} = 2 \|U-U'\|_{\HS},
    \end{align}
    where we used: the triangle inequality; matrix H{\"o}lder; and  finally, repeatedly used the basic matrix inequality $\|AB\|_{\HS} \leq \|A\|_{\oper} \cdot \|B\|_{\HS}$, together with $\|\Sigma\|_{\oper} = \delta = 1 = \|U\|_{\oper} = \|V\|_{\oper}$ and $\|\sigma^{-1}\|_{\HS} = 1$.
\end{proof}

We can now easily obtain the following corollary:
\begin{theorem}[Certification lower bound for nearly-mixed states]
\label{thm:almost-mixedness-testing2}
    Let $d \geq 2$, let $0 < \eps \leq 1/2$, and let $\sigma \in \mathbb{C}^{d \times d}$ be a quantum state with all eigenvalues at least $2\eps/d$. 
    Then any algorithm that $\eps$-certifies~$\sigma$ must use 
    \begin{equation}
        \Omega\left(\frac{d^{5/2}}{\|\sigma^{-1}\|_{\HS} \cdot \eps^2}\right)
    \end{equation}
    copies of the unknown state.  In particular, this is $\Omega(d/\eps^2)$ if  all eigenvalues of~$\sigma$ are~$\Omega(1/d)$.
\end{theorem}
\begin{proof}
    Define $\delta = 2\eps/d$. 
    Let $\Sigma = \Sigma(d,\delta) \in \mbb{C}^{d \times d}$ denote $\diag(-\delta, +\delta, \dots, -\delta, +\delta)$ if $d$ is even; if $d$ is odd, let the last diagonal entry of~$\Sigma$ be~$0$ instead.
    For $U \in U(d)$, define 
    \begin{equation}
        \rho_U \triangleq \sigma + \Delta_U, \quad\text{where}\quad \Delta_U \triangleq U \Sigma U^\dag.
    \end{equation}
    Here $\rho_U$ is a valid quantum state because $\Sigma$ is traceless and has operator norm at most the smallest eigenvalue of~$\sigma$. 
    Moreover $\|\sigma - \rho_U\|_1 = \|\Delta_U\|_1 \geq (d-1)\delta = \frac{d-1}{d} \cdot 2\eps \geq \eps$.
    Thus a certification algorithm for~$\sigma$ must be able to distinguish $\rho_{\bfu}^{\otimes n}$ from $\sigma^{\otimes n}$ with high probability, when $\bfu \sim U(d)$  is drawn according to Haar measure.  But combining \Cref{lem:quantum-ingster-suslina,lem:exp-bound}, we have that
    \begin{equation}
        n \leq .02 \frac{\sqrt{d}} {\|\sigma^{-1}\|_{\HS} \cdot \delta^2} = .005 \frac{d^{5/2}} {\|\sigma^{-1}\|_{\HS} \cdot \eps^2} \quad\implies\quad
        \qdivchi \Bigl(\mbb{E}_{\bfu} \rho_{\bfu}^{\otimes n} \bigm\| \sigma^{\otimes n}\Bigr) \leq .02,
    \end{equation}
    completing the proof of the lower bound.
\end{proof}

\subsection{Generalizing to Bucketed States} \label{sec:warm-up-3}
As discussed in \Cref{sec:tech-lower-us}, the preceding
\Cref{thm:almost-mixedness-testing2}
does not give a good bound if $\sigma$ has very small eigenvalues.
For example, suppose $\sigma$ essentially looks like the maximally mixed state on a subspace of dimension $r < d$; say, it has $\Omega(r)$ eigenvalues that are $\Omega(1/r)$, and all other eigenvalues are extremely tiny. Then the correct certification lower bound should be $\Omega(r/\eps^2)$. However, $\sigma$'s tiny eigenvalues could make $\|\sigma^{-1}\|_{\HS}$ very large, rendering  \Cref{thm:almost-mixedness-testing2} useless. 

In this case, the solution is to make a lower bound instance in which only the $\Omega(1/r)$-large eigenspace of~$\sigma$ is perturbed.
More generally, one can bucket the eigenvalues of~$\sigma$ into roughly-equal buckets, forming a block-diagonal decomposition; then, one can make independent perturbations on each block.
Here we work out what results from a general bucketing:
\begin{lemma}
\label{lem:perturbation-dependent-quantum-paninski0}
    Let $\sigma \in \mbb{C}^{d \times d}$ be a quantum state with eigenvalues $\lambda_1, \dots, \lambda_d$.
    Let $S_1, \dots, S_{\ell}$ be a collection of disjoint subsets of~$[d]$ satisfying:
    \begin{itemize}
        \item $d_j \triangleq |S_j| \geq 2$ for each $j \in [\ell]$;
        \item for each $j \in [\ell]$, there is a value $\underline{\lambda}_j > 0$ such that $\lambda_i \in (\underline{\lambda}_j, 2\underline{\lambda}_j]$ for all $i \in S_j$. 
    \end{itemize}
    
    Choose $0 \leq \eps_j \leq \underline{\lambda}_j$ for each $j \in [\ell]$, and suppose $\sum_j \eps_j \cdot 2 \lfloor d_j/2 \rfloor \geq \eps$.
    Then any algorithm that $\eps$-certifies~$\sigma$ must use 
    \begin{equation}
        n = \Omega\left(
            \sum_{j = 1}^{\ell} \epsilon_j^4/\underline{\lambda}_{j}^{2}\right)^{-1/2}
    \end{equation}
    copies of the unknown state.
\end{lemma}

\begin{proof}
    Let $S_{0} = [d] \setminus \bigcup_{j \geq 1} S_j$.
    We may assume without loss of generality that $\sigma$ has a block-diagonal structure,
    $
        \sigma = \bigoplus_{j=0}^\ell \sigma_j,
    $ 
    where the eigenvalues of $\sigma_j$ are those in~$S_j$.  We observe that
    \begin{equation}
    \label{eq:bucketing-hsnorm}
        \|\sigma_j^{-1}\|_2^2 = \sum_{i \in S_j} \lambda_i^{-2} \leq d_j/\underline{\lambda}_j^2,
    \end{equation}
    as $\sigma_j$'s eigenvalues exceed $\underline{\lambda}_j$. 
    Recalling the diagonal matrices $\Sigma(d,\delta)$ from the proof of \Cref{thm:almost-mixedness-testing2}, we define $\Sigma_j \triangleq \Sigma(d_j, \eps_j)$ for $1 \leq j \leq \ell$.
    Then given a sequence $\vec{U} = (U_1, \dots, U_\ell)$, where $U_j \in U(d_j)$, we define $\Delta_j = U_j \Sigma_j U_j^\dag$, and 
    \begin{equation}
        \rho_{\vec{U}} \triangleq \sigma_0 \oplus \bigoplus_{j=1}^\ell \rho_{j, U_j}, \quad\text{where}\quad \rho_{j,U_j} \triangleq \sigma_j + \Delta_j;
    \end{equation}
    similar to the proof of \Cref{thm:almost-mixedness-testing2}, this is a valid quantum state satisfying 
    \begin{equation}
        \|\sigma - \rho_{\vec{U}}\|_1 = \sum_{j=1}^\ell \|\Delta_j\|_1 = \sum_{j=1}^\ell \eps_j \cdot 2\lfloor d_j/2\rfloor \geq \eps.
    \end{equation}
    Thus our lower bound may be concerned with distinguishing $\rho_{\vec{\bfu}}^{\otimes n}$ from $\sigma^{\otimes n}$, where each $\bfu_j \sim U(d_j)$ independently according to Haar measure.
    By \Cref{lem:quantum-ingster-suslina}, we have
    \begin{align}
        1 + \qdivchi \Bigl(\mbb{E}_{\vec{\bfu}} \rho_{\vec{\bfu}}^{\otimes n} \bigm\| \sigma^{\otimes n}\Bigr) &\leq \mbb{E}_{\vec{\bfu},\vec{\bfu}'} \exp\left(n \cdot \sum_{j=1}^\ell \tr(\sigma_j^{-1} \bfDelta_j \bfDelta_j^\prime)\right)\\
        &= \prod_{j=1}^\ell \mbb{E}_{\vec{\bfu},\vec{\bfu}'} \exp\left(n \cdot \tr(\sigma_j^{-1} \bfDelta_j \bfDelta_j^\prime)\right) \\
        &\leq \prod_{j=1}^\ell \exp\left(\frac{48n^2 \cdot \|\sigma_j^{-1}\|_2^2 \cdot \eps_j^4}{d_j}\right) 
        = \exp\left(48n^2 \sum_{j=1}^\ell \frac{\|\sigma_j^{-1}\|_2^2 \cdot \eps_j^4}{d_j}\right) \label{eqn:combine}
    \end{align}
    Above, the first equality used independence of the $\bfu_j$'s and $\bfu_j^\prime$'s; the subsequent inequality used \Cref{lem:exp-bound}.
    Finally, using \Cref{eq:bucketing-hsnorm}, we get \begin{equation}
        \sum_{j=1}^\ell \frac{\|\sigma_j^{-1}\|_2^2 \cdot \eps_j^4}{d_j} 
        \leq
        \sum_{j=1}^\ell \eps_j^4/\underline{\lambda}_j^2.
    \end{equation}
    Combining this with \Cref{eqn:combine} and using $\exp(x) - 1 \sim x$ for small~$x$ shows  $\qdivchi \Bigl(\mbb{E}_{\vec{\bfu}} \rho_{\vec{\bfu}}^{\otimes n} \bigm\| \sigma^{\otimes n}\Bigr) = o(1)$ unless $n = \Omega\left(
            \sum_{j = 1}^{\ell} \epsilon_j^4/\underline{\lambda}_{j}^{2}\right)^{-1/2}$, as needed.
\end{proof}

\section{Lower Bound}
\label{sec:lower}
In this section, we will prove our main lower bound for state certification with entangled measurements, stated in the following theorem:
\begin{theorem}[Nearly Instance-Optimal Lower Bound]
    \label{thm:lower-main}
     Fix $0 < \epsilon < 1/12$. Let $\sigma \in \mathbb{C}^{d \times d}$ be a density matrix. Let $\sigma^*$ be the variant of~$\sigma$ given by zeroing out at most $12 \epsilon$ eigenvalue mass from $\sigma$ (see \Cref{def:bucketing-scheme-lower-bound} for details on mass removal). Then, any algorithm that $\eps$-certifies~$\sigma$ must use at least $\Tilde{\Omega}\left( \|\sigma^*\|_{1/2}/\epsilon^2\right)$ copies of the unknown state.
\end{theorem}

We start by defining a bucketing and mass removal scheme that splits~$\sigma$ into blocks that are ``nearly-maximally mixed'' (in the sense of \Cref{thm:almost-mixedness-testing2}) in \Cref{sec:bucketing-lower}. Then, we construct a lower bound instance using independent perturbations to each such block and use it to show our (nearly final) lower bound in \Cref{sec:lower-block-diag}. Finally, this bound is not strong enough in a certain corner case: when $\sigma$ has a largest eigenvalue near~$1$. We handle this corner case in \Cref{sec:lower-corner}. Finally, we put these two bounds together to conclude the proof of \Cref{thm:lower-main} in \Cref{sec:lower-final}.
\subsection{Bucketing and Mass Removal}
\label{sec:bucketing-lower}
In this section, we present the bucketing and mass removal scheme which will be used throughout \Cref{sec:lower}. It is similar to, but not quite the same as, the one in~\cite[Definition 5.2]{chen2022toward}. In particular, our scheme differs in the order in which we sort eigenvalues as well as the amount of mass we remove.

\begin{definition}[Bucketing and Mass Removal Scheme for Lower Bound]
\label{def:bucketing-scheme-lower-bound}
    Let $\sigma \in \mbb{C}^{d \times d} = \diag(\lambda_1, \dots ,\lambda_d)$ be a quantum state. (Recall: the assumption that $\sigma$ is diagonal is without loss of generality.) We will also assume that all eigenvalues of $\sigma$ are non-zero. This assumption is also without loss of generality, as one can easily restrict our arguments to the subspace corresponding to nonzero eigenvalues. 
    \begin{itemize}
        \item For $j \in \mbb{Z}_{\geq 0}$, we define the $j^{\mathrm{th}}$ bucket, $S_j \triangleq \left\{i \in [d], \lambda_i \in \left(\frac{1}{2^{j+1}}, \frac{1}{2^{j}}\right]\right\}$. Let $d_j = |S_j|$. For $i \in [d]$, let $j(i)$ denote the bucket containing the index $i$.
        \item Let $\mc{J}$ denote the set of buckets for which $S_j \neq \emptyset$. Let $\ssing$ denote the set of indices $i \in [d]$ belonging to size-1 buckets (i.e., $d_{j(i)} = 1$),  and let $\smany$ denote the set of $i \in [d]$ lying in buckets of size greater than 1 (i.e., $d_{j(i)} > 1$).
        \item We assume, without loss of generality, that $\lambda_1, \dots, \lambda_d$ are sorted in increasing order of $\widehat{\lambda}_i \triangleq \lambda_i/d_{j(i)}$ with ties broken arbitrarily. Let $d^\prime < d$ be the largest index such that $\sum_{i = 1}^{d^\prime} \lambda_i \leq 12\epsilon$. Let $\stail \triangleq [d^\prime]$.
        \item Let $\imax$ denote the index of the largest eigenvalue of $\sigma$, and let $\srem \triangleq \stail \cup \{\imax\}$ denote the ``peripheral'' entries. 
        \item Let $\sigma^*$ denote the (diagonal) matrix obtained by zeroing out the entries of~$\sigma$ indexed by $\stail$. Let $\mc{J}^*$ denote the set of buckets $j \in \mc{J}$ for which $S_j$ has nonempty intersection with $[d] \backslash \stail$.
        \item Let $\sigma^\prime$ denote the matrix obtained by further zeroing out the largest entry of $\sigma^*$. Let $\mc{J}^\prime$ denote the set of buckets $j \in \mc{J}$ for which $S_j$ has nonempty intersection with $[d] \backslash (\{\imax\} \cup \stail)$.
    \end{itemize}
\end{definition}
Note that we can write $\sigma = \bigoplus_{j \in \mc{J}} \sigma_j$, where $\sigma_j$ is the principal submatrix of $\sigma$ associated with the indices in $S_j$.
Our bucketing scheme also has the following important property.

\begin{fact}[A variant of Fact 5.3 in \cite{chen2022toward}]
\label{fact:num-buckets}
    There are at most $\bigo(\mathrm{log}(d/\epsilon))$ buckets $j \in \mc{J}$ for which $S_j$ and $\stail$ are disjoint. 
\end{fact}
\begin{proof}
    Fix any $i_1 \not \in \stail$. 
    We will show $\lambda_{i_1} \geq \eps/d^2$.
    It will then follow that the only buckets that can be disjoint from $\stail$ are $S_0, S_1, \dots, S_{\lceil\log_2(d^2/\eps)\rceil}$.

    For any $i_2 \in \stail$, we have $i_2 < i_1$, hence $\widehat{\lambda}_{i_2} \leq \widehat{\lambda}_{i_1}$; i.e., $\lambda_{i_1}/d_{j(i_1)} \geq \lambda_{i_2}/d_{j(i_2)}$ and hence $\lambda_{i_1} \geq \lambda_{i_2}/d$. 
    Summing over $i_2 \in \stail$ yields $d' \cdot \lambda_{i_1} \geq \tau/d$, where $\tau \triangleq \sum_{i_2 \in \stail} \lambda_{i_2}$ (and recall $d' = |\stail|$). If $\tau \geq \eps$, then the desired $\lambda_{i_1} \geq \eps/d^2$ easily follows.
    Otherwise, $\tau < \eps$.  Then by maximality of $d'$, we have $\lambda_{d'+1} > 11\eps$, hence $\widehat{\lambda}_{d'+1} > 11\eps/d$.  But also $i_1 \geq d'+1$, so $\widehat{\lambda}_{i_1} \geq 11\eps/d$, whence $\lambda_{i_1} \geq 11\eps/(d d_{j(i+1)}) \geq \eps/d^2$, as desired.
\end{proof}
\subsection{Lower Bound from Block-Diagonal Perturbations}
\label{sec:lower-block-diag}

In this section, we will show the following lower bound:
\begin{lemma}
    \label{lem:sigma-prime-lower-bound0}
    Let $\sigma \in \mathbb{C}^{d \times d}$ be a quantum state. Let $0 < \eps < 1/12$. Any algorithm that $\eps$-certifies~$\sigma$ must use at least $\Tilde{\Omega}(\|\sigma^\prime\|_{1/2}/\epsilon^2)$
    copies of the unknown state, where $\sigma^\prime$ is defined in \Cref{def:bucketing-scheme-lower-bound}.
\end{lemma}
Note that this bound is weaker than the main lower bound in \Cref{thm:lower-main}, as it is based on $\sigma^\prime$ rather than $\sigma^*$ (recall that $\sigma^\prime$ is obtained by additionally zeroing out the largest eigenvalue from $\sigma^*$). We will later improve upon this bound in \Cref{sec:lower-final}.

Now to obtain \Cref{lem:sigma-prime-lower-bound0}, we hope to apply \Cref{lem:perturbation-dependent-quantum-paninski0} for bucketed states to our bucketing scheme in \Cref{def:bucketing-scheme-lower-bound}. We recall that \Cref{lem:perturbation-dependent-quantum-paninski0} requires buckets to have multiple entries (not be singletons). However, in the extreme case when all buckets only have a single entry each, we cannot meaningfully use this bound for bucketed states. We first handle such corner cases, where $\sigma^\prime$ is dominated by single-entry buckets, in the following lemma:
\begin{lemma}[Similar to Lemma 5.12 in \cite{chen2022toward}] 
\label{lem:corner-case-singleton-buckets-dominate0}
    \Cref{lem:sigma-prime-lower-bound0} holds 
    if $\sum_{i \in \ssing \setminus \srem} \lambda_i^{1/2} > \frac{1}{2} \|\sigma^\prime\|_{1/2}^{1/2}$.
\end{lemma}

We will need the following instance-optimal lower bound for testing identity of probability distributions and its corollary below.
\begin{lemma}[Lower bound from \cite{valiant2017automatic}]
    Given a known distribution $q$ and samples from an unknown distribution $p$, distinguishing between $p = q$ and $\|p-q\|_1 \geq \epsilon$ with high probability requires at least $\Omega(\|q^{-\max}_{-\epsilon}\|_{2/3}/\epsilon^2)$ samples from $p$, where $q^{-\max}_{-\epsilon}$ is obtained by zeroing out the bottom $\epsilon$ mass and the largest entry of $q$.
\end{lemma}

Note that a classical distribution testing lower bound immediately implies one for the analogous quantum testing problem (by restricting to diagonal quantum states).  This gives us the following corollary:
\begin{corollary}
\label{cor:valiant-lower-quantum}
Let $\sigma \in \mbb{C}^{d \times d}$ be a quantum state. Any algorithm that $\epsilon$-certifies $\sigma$ must use at least $\Omega(\|\sigma^{-\max}_{-\epsilon}\|_{2/3}/\epsilon^2)$ copies of the unknown state, where $\sigma^{-\max}_{-\epsilon}$ is obtained by zeroing out the largest entry and the bottom $\epsilon$ probability mass of the eigenvalues of $\sigma$.
\end{corollary}

Now, we can prove \Cref{lem:corner-case-singleton-buckets-dominate0}.

\begin{proof}[Proof of \Cref{lem:corner-case-singleton-buckets-dominate0}]
    In this case, we will see that the classical lower bound of \cite{valiant2017automatic} already suffices. From \Cref{cor:valiant-lower-quantum}, we obtain a lower bound of $\Omega(\|\sigma^{-\max}_{-\epsilon}\|_{2/3}/\epsilon^2)$. Thus it suffices to show $\|\sigma^{-\max}_{-\epsilon}\|_{2/3} \geq \Omega(\|\sigma^\prime\|_{1/2})$.  In fact, we will show
    \begin{equation}
        \|\sigma^{-\max}_{-\epsilon}\|_{2/3} \geq \lambda_{i^*} \geq \Omega(\|\sigma^\prime\|_{1/2}), \quad \text{where } i^* \triangleq \argmax\{\lambda_i : i \in \ssing \setminus \srem\}.\label{eqn:leftright}
    \end{equation}
    (Note that $\ssing \setminus \srem \neq \emptyset$ by virtue of the lemma's hypothesis.) 
    We first establish the left inequality in \Cref{eqn:leftright}.  If $\lambda_{i^*}$ is \emph{not} one of the entries that is zeroed out when $\sigma^{-\max}_{-\epsilon}$ is formed, then the inequality certainly holds.  Otherwise, $\lambda_{i^*}$ must belong to the bottom $\epsilon$ mass of~$\sigma$. Thus any nonzero entry of $\sigma^{-\max}_{-\epsilon}$ must be at least~$\lambda_{i^*}$. This verifies the left inequality in \Cref{eqn:leftright}, unless $\sigma^{-\max}_{-\epsilon} = 0$.  But if all of $\sigma^{-\max}_{-\epsilon}$'s entries are zeroed out, then $\sigma$~must have one eigenvalue at least $1-\eps$ and all other eigenvalues adding up to at most $\eps$. In this case,  $\sigma^\prime$ would also be zero, and the lemma is trivially true.

    We turn to establishing the right inequality in \Cref{eqn:leftright}. Using the lemma's hypothesis and $2^{-j(i)+1} < \lambda_i \leq 2^{-j(i)}$, we have
    \begin{align}
    \frac12 \|\sigma^\prime\|_{1/2}^{1/2} < \sum_{i \in \ssing \setminus \srem} \lambda_i^{1/2} \leq \lambda_{i^*}^{1/2} + \sum_{i \in \ssing \setminus \srem \setminus \{i^*\}} 2^{-j(i)/2} &\leq \lambda_{i^*}^{1/2} + 2^{-(j(i^*)+1)/2} + 2^{-(j(i^*)+2)/2} + \cdots \nonumber\\
        &\leq \lambda_{i^*}^{1/2} + \Theta(2^{-j(i^*)/2}) \leq \bigo(\lambda_{i^*}^{1/2}),
    \end{align}
    where in the third inequality we use the fact that eigenvalues in single-entry buckets must be in geometric progression. This confirms the right inequality in \Cref{eqn:leftright}.
\end{proof}

Given the above, to complete the proof of \Cref{lem:sigma-prime-lower-bound0}, it suffices to show the following:
\begin{lemma}
    \label{lem:tuning-the-perturbations0}
     \Cref{lem:sigma-prime-lower-bound0} holds if $\sum_{i \in \ssing \backslash \srem} \lambda_i^{1/2} \leq \frac{1}{2} \|\sigma^\prime\|_{1/2}^{1/2}$.
\end{lemma}

We will start by applying \Cref{lem:perturbation-dependent-quantum-paninski0} to the bucketing scheme from \Cref{def:bucketing-scheme-lower-bound}.

\begin{corollary}
\label{cor:perturbation-dependent-bound}
    Let $0 < \eps < 1/12$. Let $\sigma \in \mbb{C}^{d \times d}$ be a quantum state, with buckets $\{S_j\}_{j \in \mc{J}}$ as in \Cref{def:bucketing-scheme-lower-bound}. For $j \in \mc{J}$ with $d_j > 1$, let $0 \leq \epsilon_j \leq 2^{-j-1}$ such that $\sum_{j} \epsilon_j \cdot 2 \lfloor d_j/2 \rfloor \geq \epsilon$. Then any algorithm that $\eps$-certifies $\sigma$ must use 
    \begin{equation}
        n \geq \Omega\left(
            \sum_{j \in \mc{J}, d_j > 1} \epsilon_j^4 2^{2j}\right)^{-1/2}
    \end{equation}
    copies of the unknown state.
\end{corollary}
\begin{proof}
    Directly apply \Cref{lem:perturbation-dependent-quantum-paninski0} to the buckets $S_j$ with $d_j > 1$, and with $\underline{\lambda}_j = 2^{-j-1}$. 
\end{proof}

Now, we will appropriately tune the perturbations $\epsilon_j$ in \Cref{cor:perturbation-dependent-bound} to prove \Cref{lem:tuning-the-perturbations0}.

\begin{proof}[Proof of \Cref{lem:tuning-the-perturbations0}.]
    For $j \in \mc{J}$ with $d_j > 1$, let $\epsilon_j = \min\{2^{-j-1}, \alpha d_j^{1/3} 2^{-2(j+1)/3}\}$. 
    Here, $\alpha$ is a normalizing factor such that
    \begin{equation}
        \sum_{j \in \mc{J}, d_j > 1} 2 \lfloor d_j/2 \rfloor \epsilon_j = \epsilon.
    \end{equation}
    We will analyze $\alpha$ in \Cref{lem:normalizing-factor-upper-bound0} below.
    Recall that from \Cref{cor:perturbation-dependent-bound}, we have $n \geq \Omega\left(
                \sum_j \epsilon_j^4 2^{2j}
        \right)^{-1/2}.$
    We may upper-bound the quantity in parentheses as follows:
    \begin{align}
        \sum_{j \in \mc{J}, d_j > 1} \epsilon_j^4 2^{2j}
        &\leq \bigo\left(\sum_{j \in \mc{J}, d_j > 1} \min\{ 2^{-2j}, \alpha^4 d_j^{4/3} 2^{-2j/3}\}\right)\\
        &\leq \bigo\left(\sum_{j \in \mc{J}, d_j > 1} \min\{ \alpha^3 d_j 2^{-j}, \alpha^4 d_j^{4/3} 2^{-2j/3}\}\right) \tag{as $\min\{x,y\} \leq \{x^{1/4} y^{3/4}, y\}$}\\
        &= \alpha^3 \cdot \bigo\left(\sum_{j \in \mc{J}, d_j > 1} d_j \cdot \min\{  2^{-j}, \alpha d_j^{1/3} 2^{-2j/3}\}\right) \\
        &= \alpha^3 \cdot \bigo\left(\sum_{j \in \mc{J}, d_j > 1} d_j \eps_j\right) = \bigo(\alpha^3 \eps).
    \end{align}
    Thus $n \geq \Omega(\alpha^{-3/2} \eps^{-1/2})$.  Now in 
    \Cref{lem:normalizing-factor-upper-bound0} below we will show:
    \begin{equation}
        \alpha \leq \bigo(\epsilon) \cdot \left( \sum_{j \in \mc{J}^\prime, d_j > 1} \max_{i \in S_j} \{\lambda_i^{2/3} d_j^{4/3}\}\right)^{-1}.
    \end{equation}    
    We thereby conclude:
    \begin{align}
        n \eps^2 &\geq \Omega\left( \sum_{j \in \mc{J}^\prime, d_j > 1} \max_{i \in S_j} \{\lambda_i^{2/3} d_j^{4/3}\}\right)^{3/2} 
        \geq \Omega\left( \max_{j \in \mc{J}^\prime, d_j > 1} \max_{i \in S_j} \{\lambda_i d_j^{2}\}\right)\\
        &\geq \Omega\left(\frac{1}{|\mc{J}^\prime|} \sum_{j \in \mc{J}^\prime, d_j > 1} \max_{i \in S_j} \{\lambda_i^{1/2} d_j\}\right)^2
        \geq \Omega \left(\frac{1}{|\mc{J}^\prime|^2 }\right) \left(
            \sum_{j \in \mc{J}^\prime, d_j > 1, i \in S_j} \lambda_i^{1/2}
        \right)^2 \geq\Omega\left(\frac{\|\sigma^\prime\|_{1/2}}{\log(d/\epsilon)^2}\right),
    \end{align}
    where in the last inequality we use \Cref{fact:num-buckets} and the hypothesis of the lemma.
    This completes the proof.
    
\end{proof}

It remains to upper-bound the normalizing factor $\alpha$.
\begin{lemma}
\label{lem:normalizing-factor-upper-bound0}
For $j \in \mc{J}, $ let $\epsilon_j = \min\{2^{-j-1}, \alpha d_j^{1/3} 2^{-2(j+1)/3}\}$ if $d_j > 1$ and $\epsilon_j = 0$ if $d_j = 1$, where $\alpha$ is chosen so that $\sum_{j \in \mc{J}, d_j > 1} 2 \lfloor d_j/2 \rfloor \epsilon_j = \epsilon$. Then, under the hypothesis of \Cref{lem:tuning-the-perturbations0},
    \begin{equation}
        \alpha \leq \bigo(\epsilon) \cdot \left( \sum_{j \in \mc{J}^\prime, d_j > 1} \max_{i \in S_j} \{\lambda_i^{2/3} d_j^{4/3}\}\right)^{-1}.
    \end{equation}
\end{lemma}

Before proving the above lemma, we first state the following elementary fact from \cite{chen2022toward}.

\begin{fact}[Fact 5.16 in \cite{chen2022toward}]
\label{fact:sums-of-sorted-sequences}
     Let $0 < \epsilon^\prime < 1$. Let $u_1 < \dots < u_m$ and $v_1 \leq \dots \leq v_n$ be numbers with $u_{i+1} \geq 2u_i$ for all $i$. Let $w_1 \leq \dots \leq w_{m+n}$ be these numbers in sorted order. Let $d_1, \dots, d_n$ be arbitrary integers. For $i \in [m+n]$, define $d_i^*$ to be 1 if $w_i$ corresponds to some $u_j$, and $d_i^* = d_j$ if $w_i$ corresponds to some $v_j$.

    Let $s$ be the largest index for which $\sum_{i = 1}^s w_i d_i^* \leq 3\epsilon^\prime$. Let $a,b$ be the largest indices for which $u_a, v_b$ are present among $w_1, \dots, w_s$ (if none exists, take it to be $0$). Then either $b = n$ or $\sum_{i = 1}^{b+1} v_id_i > \epsilon^\prime$.
\end{fact}

\Cref{fact:sums-of-sorted-sequences} allows us to prove the following lemma relating the two possible choices of $\epsilon_j$, similarly to Corollary 5.17 of \cite{chen2022toward}.

\begin{lemma} 
\label{lem:perturbation-choices-relation0}
If $\sum_{i \in \ssing \backslash \srem} \lambda_i^{1/2} < \frac12 \|\sigma^\prime\|_{1/2}^{1/2}$, then $\smany \backslash \srem$ is nonempty, and for any $j \in \mc{J}^\prime$ with $d_j > 1$, it holds that $\alpha \cdot 2^{-2(j+1)/3}d_j^{1/3} \leq 2^{-j-1}$.
\end{lemma}

\begin{proof}
The hypothesis of the lemma can be restated as $\sum_{i \in \smany \backslash \srem} \lambda_i^{1/2} \geq \frac12 \|\sigma^\prime\|_{1/2}^{1/2}$, which directly implies the first part of the lemma.

For the second part, assume for the sake of contradiction that for some $i^* \in \smany \backslash \srem$, which lies in some bucket $j^* \in \mc{J}^\prime$, we have that $2^{-j^*-1} < \alpha \cdot 2^{-2(j^*+1)/3}d_{j^*}^{1/3}$, or equivalently, $(\lambda_{i^*}/d_{j^*})^{1/3} \leq (2^{-j^*}/d_{j^*})^{1/3} < 2^{1/3} \alpha$.
Recall that in \Cref{def:bucketing-scheme-lower-bound}, we sort the eigenvalues in increasing order of $\lambda_i/d_{j(i)}$. As a result, for any $i \leq i^*$, we also have that $2^{-j(i)-1} < 2^{1/3}\alpha \cdot 2^{-2(j(i)+1)/3} d_{j(i)}^{1/3}$. Thus, we have that
\begin{align}
    \epsilon 
    &= \sum_{j \in \mc{J}, d_j > 1} 2 \lfloor d_j/2 \rfloor \cdot \min \{2^{-j-1}, \alpha \cdot 2^{-2(j+1)/3}d_j^{1/3}\}
    \\&\geq \sum_{j \in \mc{J}, j \geq j^*, d_j > 1} 2^{-1/3} 2 \lfloor d_j/2 \rfloor 2^{-j-1}
    \\& \geq \sum_{i \in \smany, i \leq i^*} \lambda_i/4
    \\& > \epsilon,
\end{align}
where in the third line we used that $2^{-j(i)} \geq \lambda_i$, $2 \lfloor d_j/2 \rfloor \geq 2d_j/3$ for all $d_j > 1$, and that $3 \times 2^{1/3} < 4$. In the last line, we apply \Cref{fact:sums-of-sorted-sequences} to the numbers $\{u_i\} \triangleq \{\lambda_i\}_{i \in \ssing}, \{v_i\} \triangleq \{\lambda_i/d_{j(i)}\}_{i \in \smany, i \leq i^*},\{d_i\} \triangleq \{d_{j(i)}\}_{i \in \smany, i \leq i^*},$ and $\epsilon^\prime = 4\epsilon$, giving us a contradiction.
\end{proof}
We can finally prove \Cref{lem:normalizing-factor-upper-bound0}.
\begin{proof}[ Proof of \Cref{lem:normalizing-factor-upper-bound0}]
    We have
    \begin{align}
        \epsilon &= \sum_{j \in \mc{J}, d_j > 1} 2 \lfloor d_j/2 \rfloor \min \{2^{-j-1}, \alpha \cdot d_j^{1/3} 2^{-2(j+1)/3}\}
        \\&\geq \sum_{j \in \mc{J}^\prime, d_j > 1} \alpha \cdot 2 \lfloor d_j/2 \rfloor 2^{-2(j+1)/3} d_j^{1/3}
        \\& \geq \Omega(\alpha) \sum_{j \in \mc{J}^\prime, d_j > 1} \max_{i \in S_j} \{\lambda_i^{2/3} d_j^{4/3}\}, 
    \end{align}
    where in the second line we use \Cref{lem:perturbation-choices-relation0}, and in the last line we use $\lambda_i \leq 2^{-j(i)}$, thus giving us the desired upper bound on $\alpha$.
\end{proof}

\subsection{Corner Case}
\label{sec:lower-corner}
In this section, we present an $\Omega(1/\epsilon^2)$ lower bound for state certification when the largest eigenvalue of the hypothesis state is at least $1/2$. This lets us resolve a corner case in our final bound. 
\begin{lemma}
\label{lem:epsilon2-lower-bound}
    Let $\sigma = \diag(\lambda_1, \dots, \lambda_d)$ be a quantum state, with the eigenvalues in descending order $\lambda_1 \geq \dots \geq \lambda_d$, and $\lambda_1 \geq 1/2$. Let $0 < \epsilon < \frac{1}{2\sqrt{2}}$. Then, any algorithm that $\eps$-certifies~$\sigma$ must use at least $\Omega(1/\epsilon^2)$ copies of the unknown state.   
\end{lemma}

We will divide this bound into two cases, when $\lambda_2 \leq 1/4$ and when $\lambda_2 > 1/4$, using different instances for each case. 
\begin{lemma}
    \label{lem:epsilon2-lower-bound-case1}
    \Cref{lem:epsilon2-lower-bound} holds provided $\lambda_1 \geq 1/2$ and 
    $\lambda_2 \leq 1/4$.
\end{lemma}
\begin{proof}
    For this case, we consider an instance based on small rotations to the first two eigenvectors of $\sigma$.  
    Let us write
    \begin{equation}
        \overline{\sigma} \triangleq \diag(\lambda_1, \lambda_2) = \lambda_1 \ketbra{1}{1} + \lambda_2 \ketbra{2}{2}, \qquad \sigma_{\text{tail}} \triangleq \diag(\lambda_3, \dots, \lambda_d), \qquad \text{hence } \sigma = \overline{\sigma} \oplus \sigma_{\text{tail}}.
    \end{equation}
    Also, let $R_{2\eps}$ denote a qubit rotation operator on $\mathbb{C}^2$ such that,  writing $\ket{1^\prime} = R_{2\eps} \ket{1}$ and $\ket{2^\prime} = R_{2\eps} \ket{2}$, we have $|\braket{1}{2^\prime}| = |\braket{2}{1^\prime}| = 2\eps$. 
    Then define
    \begin{equation}
        \overline{\sigma}^\prime = R_{2\eps} \overline{\sigma} R_{2\eps}^\dag = \lambda_1 \ketbra{1^\prime}{1^\prime} + \lambda_2 \ketbra{2^\prime}{2^\prime}.
    \end{equation}
    
    We will consider distinguishing $\sigma$ from $\sigma^\prime$, defined by
    \begin{equation}
        \sigma^\prime \triangleq \overline{\sigma}^\prime \oplus \sigma_{\text{tail}}.
    \end{equation}
    
    First, we show that $\sigma$ and $\sigma^\prime$ are at least $\epsilon$-far with respect to $1$-norm:
    \begin{align}
    \|\sigma - \sigma^\prime\|_1 = \|\overline{\sigma} - \overline{\sigma}^\prime\|_1&= 
    \left\|
    \lambda_1 (\ketbra{1}{1} - \ketbra{1^\prime}{1^\prime}) + \lambda_2 (\ketbra{2}{2} - \ketbra{2^\prime}{2^\prime})
    \right\|_1
    \\&\geq \lambda_1 \left\|\ketbra{1}{1} - \ketbra{1^\prime}{1^\prime}\right\|_1 - \lambda_2 \left\|\ketbra{2}{2} - \ketbra{2^\prime}{2^\prime}\right\|_1
    \\&= (\lambda_1 - \lambda_2) \left\|\ketbra{1}{1} - \ketbra{1^\prime}{1^\prime}\right\|_1
    \\&\geq \frac{1}{2} \dtr(\ketbra{1}{1}, \ketbra{1^\prime}{1^\prime}) 
    \\&= \frac12\sqrt{1 - |\braket{1}{1^\prime}|^2} = \frac12|\braket{1}{2'}| = \epsilon,
    \end{align}
    where in the second line we use the triangle inequality, in third line we use $\ketbra{1}{1} + \ketbra{2}{2} = \ketbra{1^\prime}{1^\prime} + \ketbra{2^\prime}{2^\prime}$, the second-to-last line follows by our assumptions $\lambda_1 \geq 1/2$ and $\lambda_2 \leq 1/4$, and the last line uses a standard formula for the trace distance between two pure states.
    Next, we will lower-bound the fidelity between $\sigma, \sigma^\prime$; precisely, we wish to show
    \begin{equation} \label{eqn:toshow}
        F(\sigma, \sigma^\prime) \geq (1-4\epsilon^2)^2.
    \end{equation}
    We have 
    \begin{multline}\label{eqn:form1}
        \sqrt{F(\sigma, \sigma^\prime)} = \|\sqrt{\sigma} \sqrt{\sigma^\prime}\|_1 = 
        \left\|(\sqrt{\overline{\sigma}} \oplus \sqrt{\sigma_{\text{tail}}}) \cdot (\sqrt{\overline{\sigma}^\prime} \oplus \sqrt{\sigma_{\text{tail}}})\right\|_1 = \left\|\sqrt{\overline{\sigma}}\sqrt{\overline{\sigma}^\prime} \oplus \sigma_{\text{tail}}\right\|_1 \\= \|\sqrt{\overline{\sigma}}\sqrt{\overline{\sigma}^\prime}\|_1 + \tr(\sigma_{\text{tail}})
        = \tau \sqrt{F(\sigma_2,\sigma_2^\prime)} + 1-\tau,
    \end{multline}
    where 
    \begin{equation}
        \tau \triangleq \lambda_1 + \lambda_2, \qquad \sigma_2 = \frac{\overline{\sigma}}{\tau}, \qquad \sigma_2^\prime = \frac{\overline{\sigma}^\prime}{\tau}.
    \end{equation}
    Thus it remains to investigate $F(\sigma_2, \sigma_2^\prime)$.
    Since
    \begin{equation}
        \sigma_2 = \frac{\lambda_1}{\tau} \ketbra{1}{1} + \frac{\lambda_2}{\tau} \ketbra{2}{2} = \frac12 \left(\mathbbm{1} + \alpha \ketbra{1}{1}\right), \qquad \text{for } \alpha \triangleq \frac{\lambda_1-\lambda_2}{\lambda_1 + \lambda_2},
    \end{equation}
    the Bloch vector representation of $\sigma_2$ is $\vec{a} = (0, 0, \alpha)$.
    Then it is not hard to show that the Bloch vector representation of $\sigma_2^\prime = R_{2\eps} \sigma_2 R_{2\eps}^\prime$ is 
    \begin{equation}
        \vec{a}^\prime = \alpha(4\epsilon\sqrt{1-4\epsilon^2}, 0, 1-8\epsilon^2).
    \end{equation}
    We can now directly compute fidelity using \Cref{lem:fid-jozsa}:
    \begin{equation}
        F(\sigma_2, \sigma_2^\prime) = \frac12(1+\vec{a} \cdot \vec{a}' +\sqrt{(1-\vec{a}\cdot \vec{a})(1-\vec{a}'\cdot \vec{a}')}) = \frac12(1+\alpha^2(1-8\epsilon^2) + (1-\alpha^2)) = 1-4\epsilon^2\alpha^2.
    \end{equation}
    We conclude from \Cref{eqn:form1} that
    \begin{equation}
        \sqrt{F(\sigma,\sigma')} = \tau \sqrt{1-4\epsilon^2\alpha^2} + 1- \tau = \sqrt{(\lambda_1+\lambda_2)^2-4\epsilon^2(\lambda_1-\lambda_2)^2} + 1 - \lambda_1 - \lambda_2.
    \end{equation}
    
    Recall we aim to show that $F(\sigma,\sigma^\prime) \geq (1 - 4\epsilon^2)^2$. To this end, it suffices to show that
    \begin{equation}
         \sqrt{(\lambda_1+\lambda_2)^2-4\epsilon^2(\lambda_1-\lambda_2)^2} \geq \lambda_1 + \lambda_2 - 4\epsilon^2.
    \end{equation} 
    Note that since $\epsilon < \frac{1}{2\sqrt{2}}$ and $\lambda_1 \geq 1/2$, the right-hand side is always positive. Taking the square and simplifying, it suffices to show that 
    \begin{align}
        \label{eq:corner-case-eqn-1}
        4\epsilon^2 &\leq 2(\lambda_1+\lambda_2) - (\lambda_1-\lambda_2)^2.
    \end{align}
    Now the left-hand side above is at most $1/2$ (since $\eps < \frac{1}{2\sqrt{2}}$), and the right-hand side is at least $3/4$ (using $\lambda_1 \geq 1/2$ and $\lambda_2 \leq 1/4$); thus \eqref{eq:corner-case-eqn-1} indeed holds. We have thus shown that $F(\sigma,\sigma^\prime) \geq (1 - 4\epsilon^2)^2$. Then, from \Cref{lem:dtr-infidelity}, we have
    \begin{align}
        \dtr(\sigma^{\otimes n}, \sigma^{\prime \otimes n}) &\leq \sqrt{(1 - F(\sigma^{\otimes n}, \sigma^{\prime \otimes n})}
        \\&= \sqrt{1 - F(\sigma, \sigma^{\prime})^{n}}
        \\&\leq \sqrt{1 - (1-4\epsilon^2)^{2n}}
        \\&\leq \sqrt{8n\epsilon^2},
    \end{align}
    where we use \Cref{lem:fidelity-tensorization} in the second line. Thus, $\dtr(\sigma^{\otimes n},\sigma^{\prime \otimes n})$ is $o(1)$ unless $n = \Omega(1/\epsilon^2)$.
\end{proof}

\begin{lemma}
    \label{lem:epsilon2-lower-bound-case2}
    \Cref{lem:epsilon2-lower-bound} holds provided 
    $\lambda_2 > 1/4$.
\end{lemma}
\begin{proof}
    For $u \in \{-1,+1\}$, define
    \begin{equation}
        \tau_u \triangleq \begin{pmatrix}-\epsilon^2 & u\epsilon/2 \\ u\epsilon/2 & \epsilon^2 \end{pmatrix}, \qquad 
        \sigma_u \triangleq \sigma + \tau_u \oplus \mathbf{0}_{d-2},
    \end{equation}
    where $\mathbf{0}_{d-2}$ denotes the $(d-2)\times(d-2)$ zero matrix.
    Note that $\tau_u$ is traceless and has eigenvalues $\pm \sqrt{1/4 + \eps^2} \cdot \eps \geq -1/4 > -\lambda_2$, where we used $\eps < \frac{1}{2\sqrt{2}}$.  Thus $\sigma_u$ is a valid quantum state.
    Moreover, $\|\sigma-\sigma_u\|_1 = \|\tau_u\|_1 = \sqrt{1+4\epsilon^2} \cdot \epsilon  \geq \epsilon$. We now use the quantum Ingster--Suslina method to establish the desired lower bound for the task of  distinguishing between $\sigma$ and $\sigma_{\bu}$, where $\bu \sim \{-1, +1\}$ is uniformly random. 
    Writing $\overline{\sigma} \triangleq \diag(\lambda_1, \lambda_2)$, \Cref{lem:quantum-ingster-suslina} tells us:
    \begin{align}
        1+ \qdivchi(\mbb{E}_{\bu}\sigma_{\bu}^{\otimes n}|| \sigma^{\otimes n}) &=1+ \qdivchi(\mbb{E}_{\bu}(\overline{\sigma} + \tau_{\bu})^{\otimes n}|| \overline{\sigma}^{\otimes n})  \\
        &\leq \mbb{E}_{\bu,\bv} \exp\left(n \cdot \tr(\overline{\sigma}^{-1} \cdot \tau_{\bu} \tau_{\bv})\right)\\   &= \mbb{E}_{\bu,\bv} \exp\left(n \cdot (\lambda_1^{-1} + \lambda_2^{-1})(\eps^4 + \bu \bv \epsilon^2/4)\right) \label{eqn:calc}\\
        &= \exp\left(n  (\lambda_1^{-1} + \lambda_2^{-1})\eps^4\right)\cosh\left(n (\lambda_1^{-1} + \lambda_2^{-1})\epsilon^2/4\right)\\
        &\leq \exp(8n  \eps^4)\cosh(2n\epsilon^2) = 1 + \bigo(n^2 \epsilon^4),
    \end{align}
    where \Cref{eqn:calc} is direct calculation, and then we used $\lambda_1^{-1} + \lambda_2^{-1} \leq 2 \lambda_2^{-1} \leq 2 \cdot 4 = 8$ (and that $\exp$ and $\cosh$ are increasing). 
    Thus, the quantum $\chi^2$-divergence is $o(1)$ unless $n = \Omega(1/\eps^2)$, as claimed.
\end{proof}
\begin{proof}[Proof of \Cref{lem:epsilon2-lower-bound}]
    The result now follows directly from \Cref{lem:epsilon2-lower-bound-case1,lem:epsilon2-lower-bound-case2}.
\end{proof}

\subsection{Instance-Optimal Lower Bound}
\label{sec:lower-final}
In this section, we will conclude the proof of \Cref{thm:lower-main}. We will make use of the following lemma about our bucketing and mass removal scheme from \cite{chen2022toward}\footnote{While \cite{chen2022toward} showed this for their specific bucketing and mass removal scheme, it is not hard to show that this lemma holds for our scheme in \Cref{def:bucketing-scheme-lower-bound} as well.}.
\begin{lemma}[From Lemma 5.26 in \cite{chen2022toward}] 
\label{lem:large-principal-eigenval0}
Fix $0 < p < 1$. Let $\sigma \in \mathbb{C}^{d \times d}$ be a diagonal density matrix. For $\sigma^\prime, \sigma^*$ defined in \Cref{def:bucketing-scheme-lower-bound}, if $\|\sigma^\prime\|_{p} = o(\|\sigma^*\|_{p})$, then $\|\sigma\|_\infty \geq 1/2$ and $\max_{j \in \mc{J}} \{d_j 2^{-pj}\} =1$.
\end{lemma}

We are now ready to prove our main lower bound, \Cref{thm:lower-main}.
\begin{proof}[Proof of \Cref{thm:lower-main}]
    First, consider the case when $\|\sigma^\prime\|_{1/2} = \Omega(\|\sigma^*\|_{1/2})$. Then, our lower bound of $\Tilde{\Omega}\left(\|\sigma^\prime\|_{1/2}/\epsilon^2\right)$ from \Cref{lem:sigma-prime-lower-bound0} already suffices to prove the theorem.

    Otherwise, we have $\|\sigma^\prime\|_{1/2} = o(\|\sigma^*\|_{1/2})$. In such cases, from the second part of \Cref{lem:large-principal-eigenval0} (with $p = 1/2$), we have that $d_j 2^{-j/2} \leq 1$ for all buckets $j \in \mc{J}^*$. Thus,
    \begin{equation}
        \|\sigma^*\|_{1/2} \leq \left(\sum_{j \in \mc{J}^*} d_j 2^{-j/2}\right)^2 \leq \mc{O}\left(\log^2(d/\epsilon)\right),
    \end{equation}
    where we use the fact that $\lambda_i \leq 2^{-j(i)}$ in the first step, and the second part of \Cref{lem:large-principal-eigenval0} and \Cref{fact:num-buckets} in the last step. Further, the first part of \Cref{lem:large-principal-eigenval0} tells us that the greatest eigenvalue of $\sigma$ is at least $1/2$, and for such cases, \Cref{lem:epsilon2-lower-bound} gives us a $\Omega(1/\epsilon^2)$ lower bound. As  $\|\sigma^*\|_{1/2} = \Tilde{\mc{O}}(1)$, we obtain the desired $\Tilde{\Omega}(\|\sigma^*\|_{1/2}/\epsilon^2)$ lower bound.
\end{proof}
\section{Upper Bound}
\label{sec:upper}
In this section, we will present an instance-dependent algorithm for state certification that nearly matches our lower bound; i.e., it is essentially instance-optimal. Our main result is presented in the following theorem:

\begin{theorem}[Nearly Instance-Optimal State Certification]
    \label{thm:upper-bound}
    There is a quantum algorithm that, given $\eps > 0$ and explicit description of a state $\sigma \in \mathbb{C}^{d \times d}$, has the following behavior:
    Given 
    \begin{equation}
        n = \Tilde{\bigo}\left( 
            \|\sigma^*\|_{1/2} /\epsilon^2
        \right) \cdot \log(1/\delta)
    \end{equation}
    copies of an unknown state $\rho \in \mathbb{C}^{d \times d}$, it distinguishes with success probability at least $1-\delta$ between the cases $\rho = \sigma$ and $\|\rho - \sigma\|_1 > \eps$.  Here $\sigma^*$ denotes the alteration of~$\sigma$ in which the eigenvectors/spaces of lowest $\eps^2/20$ total mass are zeroed out.\footnote{We note a few subtle differences between the definitions of $\sigma^*$ here and in the lower bound (\Cref{thm:lower-main}). The mass removed here is at most $\eps^2/20$ instead of $12\eps$, and indices are sorted in increasing order of $\lambda_i$ instead of $\lambda_i/d_{j(i)}$.} 
\end{theorem}

Our algorithm follows very closely that of \cite{chen2022toward}, who showed a nearly instance-optimal algorithm for state certification with \emph{unentangled} measurements. 
The main difference is that we substitute the general-measurements Hilbert--Schmidt tester of \cite{buadescu2019quantum} for their unentangled-measurements Hilbert--Schmidt tester.  This necessitates adjustment of several parameters.
With apologies to the reader, this means our presentation cannot be completely self-contained, lest we copy large portions of~\cite{chen2022toward}.  We will outline the whole argument, but will point to~\cite{chen2022toward} for some proof details.

We encapsulate here the Hilbert--Schmidt testing routine from~\cite{buadescu2019quantum}:
\begin{lemma}[Theorem 1.4 in \cite{buadescu2019quantum}]
\label{lem:hs-state-cert}
    There is a quantum algorithm $\mathsf{HSCertify}_{\epsilon,\delta}(\rho, \sigma)$
    that, given $n = \bigo(\log(1/\delta)/\eps^2)$ copies each of two unknown states $\rho, \sigma \in \mathbb{C}^{d \times d}$, distinguishes with probability at least $1-\delta$ the cases of $\rho = \sigma$ and $\|\rho - \sigma\|_{\HS} > \eps$.\footnote{We may apply this lemma even for $\eps > 2$. In this case, the lemma is vacuously true: $n = 0$ copies are needed, as $\|\rho - \sigma\|_2 > 2$ is impossible.} In particular, this can be accomplished if the algorithm is given copies of unknown~$\rho$ and an explicit description of~$\sigma$.
\end{lemma}

We present the necessary bucketing and mass removal scheme in \Cref{sec:bucketing-upper}, and prove \Cref{thm:upper-bound} in \Cref{sec:upper-proof}.

\subsection{Bucketing and Mass Removal}
\label{sec:bucketing-upper}
We recall verbatim the bucketing and mass removal scheme from Section~6.2 of \cite{chen2022toward}. Without loss of generality, the hypothesis state is assumed to be diagonal, $\sigma = \diag(\lambda_1, \dots, \lambda_d)$ with $\lambda_1 \leq \dots \leq \lambda_d$. 
\begin{definition}
\label{def:bucketing-upper}
    Let $d^* \leq d$ denote the largest index for which $\sum_{i = 1}^{d^*} \lambda_i \leq \epsilon^2/20$, and let $S_{\mathrm{tail}} \triangleq [d^*]$. Let $\sigma^*$ denote the matrix given by zeroing out the diagonal entries of $\sigma$ indexed by $S_{\mathrm{tail}}$. For $j \in \mbb{Z}_{\geq 0}$, let $S_j$ denote the indices $i \notin S_{\mathrm{tail}}$ for which $\lambda_i \in (2^{-j-1}, 2^{-j}]$, and denote $d_j \triangleq |S_j|$. Let $\mc{J}$ denote the set of indices $j$ for which $S_j \neq \emptyset$. Let $m \triangleq |\mc{J}|$ denote the number of relevant buckets, which is at most $\bigo(\mathrm{log}(d/\epsilon))$ (by \cite[Fact 6.6]{chen2022toward}).
\end{definition}
We  also recall the following notation, again verbatim from~\cite{chen2022toward}: 
For $j \in \mc{J}$, let $\rho[j,j], \sigma[j,j] \in \mbb{C}^{d \times d}$ denote the psd matrices given by respectively zeroing out the entries of $\rho,\sigma$ outside of the principal submatrix indexed by $S_j$. For distinct $j, j^\prime \in \mc{J}$, let $\rho[j,j^\prime] \in \mbb{C}^{d \times d}$ denote the Hermitian matrix given by zeroing out entries of $\rho$ outside of the two non-principal submatrices indexed by $S_j$ and $S_{j^\prime}$, and $S_{j^\prime}$ and $S_j$, respectively. Lastly, let $\hat{\rho}[j,j], \hat{\sigma}[j,j], \hat{\rho}[j,j^\prime], \hat{\sigma}[j,j^\prime]$ denote the same matrices but with trace normalized to $1$. Finally, let $\rho_{\mathrm{junk}}^\diag \in \mbb{C}^{d \times d}$ be the matrix obtained by zeroing out all entries of $\rho$ except for those in the principal submatrix indexed by $\stail$, and let $\rho_{\mathrm{junk}}^{\mathrm{off}} \in \mbb{C}^{d \times d}$ be obtained by zeroing out the two principal submatrices indexed by $\stail$ and $[d] \backslash \stail$ respectively. 

Finally, we will need the following facts about psd matrices from \cite{chen2022toward}. 
\begin{lemma}[Lemma 3.15 in \cite{chen2022toward}]
\label{lem:psd-block-1norm}
For a positive semidefinite block matrix $\rho = \begin{pmatrix}
    A & B\\B^\dag & C
\end{pmatrix}$, where $A$ and $C$ are square, we have that $\tr(A)\tr(C) \geq \|B\|_1^2$. In particular, $\|B\|_1 \leq \tr(\rho)/2$. 
\end{lemma}

\begin{fact}[Fact 6.7 in~\cite{chen2022toward}]
\label{fact:distance-normalized-operators}
Given two psd matrices $\rho,\sigma$, if 
$\ryanabs{\tr(\rho) - \tr(\sigma)} \leq \epsilon/2$ and $\|\rho-\sigma\|_1 \geq \epsilon$, then,
\begin{equation}
    \|\rho/\tr(\rho) - \sigma/\tr(\sigma)\|_1 \geq \frac{\epsilon}{2 \tr(\sigma)}.
\end{equation}
\end{fact}

\subsection{Proof of Upper Bound}
\label{sec:upper-proof}
In this section, we prove our main upper bound. We start by recording a lemma, taken directly from~\cite{chen2022toward}, that summarizes the different ways in which $\rho$ can be far from~$\sigma$.  The below is taken directly from the first part of Theorem 6.1's proof in~\cite{chen2022toward}, except we omit their second case (``$\|\rho_{\mathrm{junk}}^{\mathrm{off}}\|_1 \geq \epsilon/2$''). This is because they show~\cite[Lemma~6.9]{chen2022toward} that it is covered by the first case (using \Cref{lem:psd-block-1norm}).
\begin{lemma} (Theorem 6.1 proof in~\cite{chen2022toward}, together with their Lemma~6.9.)
\label{lem:upper-cases}
    If $\|\rho-\sigma\|_1 \geq \epsilon$, then one of the following must be true:
    \begin{enumerate}
        \item $\|\rho_{\mathrm{junk}}^{\diag}\|_1 \geq .125\epsilon^2$.
        \item There exists $j \in \mc{J}$ for which $\|\rho[j,j] - \sigma[j,j]\|_1 > .05 \epsilon/m^2$.\footnote{In \cite{chen2022toward} this was $.1\eps/m^2$, but we weaken it to $.05\eps/m^2$ to help correct a minor typo/bug.}
        \item Case 2 fails, but there exist distinct $j,j^\prime \in \mc{J}$ for which $\|\rho[j,j^\prime]\|_1 > .2\epsilon/m^2$.
    \end{enumerate}
\end{lemma}

Now, as in \cite{chen2022toward}, we can separately test for each of these three cases (using confidence parameter~$\delta/3$ for each).
Testing for Case~1, $\|\rho_{\mathrm{junk}}^{\diag}\|_1 \geq .125\epsilon^2$, is easily done with $\bigo(\log(1/\delta)/\epsilon^2)$ measurements (even unentangled ones), as is shown in \cite[Lemma 6.8]{chen2022toward}.
Thus to complete the proof of our \Cref{thm:upper-bound}, 
it suffices for us to establish the following two lemmas for testing Cases~2 and~3.

\begin{lemma}
\label{lem:upper-case-3}
    Using entangled measurements, $\Tilde{\bigo}(\|\sigma^*\|_{1/2} /\epsilon^2) \cdot \log(1/\delta)$ copies suffice to test, with probability at least $1-\delta$, whether $\rho = \sigma$ or whether Case~2 of \Cref{lem:upper-cases} holds.
\end{lemma}
\begin{proof}
    This proof very closely resembles that of~\cite[Lemma~6.10]{chen2022toward}.
    We will show, for each fixed $j \in \mathcal{J}$, that $\Tilde{\bigo}(d_j^2 2^{-j}/\epsilon^2) \cdot \log(1/\delta)$ copies suffice to distinguish, with confidence at least $1-\delta/m$, between $\rho = \sigma$ and  $\|\rho[j,j] - \sigma[j,j]\|_1 > .05\epsilon/m^2$.
    The lemma then follows by observing 
    \begin{equation}
        \sum_{j \in \mc{J}} d_j^2 2^{-j} \leq \left(\sum_{j \in \mc{J}} d_j 2^{-j/2}\right)^2 \leq 2\left(\sum_{i \in [d] \setminus \stail} \lambda_i^{1/2}\right)^2  = 2\|\sigma^*\|_{1/2}.
    \end{equation}
    
    The first step of the algorithm is to estimate $\tr(\rho[j,j])$ to within additive error~$.005\eps/m^2$.  This can easily be done (with confidence $1-.5\delta/m$) using $\bigo(m^4/\eps^2)\cdot \log(m/\delta) = \Tilde{\bigo}(1/\eps^2) \cdot \log(1/\delta)$ copies of~$\rho$, simply by empirical estimation of the projector~$\Pi_j$ onto the coordinates in~$S_j$.  If the estimate differs from $\tr(\sigma[j,j])$ by more than $.005\eps/m^2$, then we are confident $\rho \neq \sigma$ and may reject. Otherwise, we may assume 
    \begin{equation} \label{eqn:closetr}
        \abs{\tr(\rho[j,j]) - \tr(\sigma[j,j])} \leq .01\eps/m^2.
    \end{equation}

    Next, if $\tr(\sigma[j,j]) \leq .02\eps/m^2$ we may accept, as this implies 
    \begin{equation}
        \|\rho[j,j] - \sigma[j,j]\|_1 \leq \|\rho[j,j]\|_1 + \|\sigma[j,j]\|_1 \leq (.02\eps/m^2 + .01\eps/m^2) + .02\eps/m^2 \leq .05 \eps/m^2;
    \end{equation}
    i.e., we are not in the case $\|\rho[j,j] - \sigma[j,j]\|_1 > .05\eps/m^2$. Thus we henceforth assume
    \begin{equation} \label{eqn:reject}
        \tr(\sigma[j,j]) \geq .02\eps/m^2 \quad\implies\quad \tr(\rho[j,j]) \geq \tr(\sigma[j,j]) - .01\eps/m^2 \geq \tfrac12 \tr(\sigma[j,j]) \geq \tfrac14 d_j 2^{-j}.
    \end{equation}

    Let us now consider $\|\hat{\rho}[j,j] - \hat{\sigma}[j,j]\|_{\HS}$.  On one hand, if $\rho = \sigma$ then this is~$0$.  On the other hand, if $\|\rho[j,j] - \sigma[j,j]\|_1 > .05\eps/m^2$, then by \Cref{eqn:closetr} and \Cref{fact:distance-normalized-operators} we get
    \begin{equation}
        \|\hat{\rho}[j,j] - \hat{\sigma}[j,j]\|_1 > \frac{.025\epsilon/m^2}{\tr(\sigma[j,j])} \geq \frac{.025\epsilon/m^2}{d_j 2^{-j}} \quad\implies\quad 
        \|\hat{\rho}[j,j] - \hat{\sigma}[j,j]\|_{\HS} > \frac{.025\epsilon/m^2}{d_j^{3/2} 2^{-j}}.
    \end{equation}
     
    Now \Cref{lem:hs-state-cert} tells us we can distinguish these two cases 
    with confidence~$.9$ (which we will later increase to $1 - .5\delta/m$) by 
    performing entangled measurements on
    \begin{equation}\label{eqn:need}
        \bigo\left(\frac{d_j^3 2^{-2j} \cdot m^4}{\eps^2}\right) = \Tilde{\bigo}\left(\frac{d_j^3 2^{-2j}}{\eps^2}\right)
    \end{equation}
    copies of $\hat{\rho}[j,j]$. We do not quite have access to copies of~$\hat{\rho}[j,j]$, but \Cref{eqn:reject} shows that by performing the projective measurement $\Pi_j$ on a copy of~$\rho$, we can get a copy of~$\hat{\rho}[j,j]$ with probability~$\Omega(d_j 2^{-j})$.  Thus in expectation, $\bigo(2^j/d_j)$ copies of~$\rho$ suffice to get $1$ copy of~$\hat{\rho}$; hence with $\Tilde{\bigo}\left(d_j^2 2^{-j}/\eps^2\right)$ copies of~$\rho$, we get can get the needed number of copies in \Cref{eqn:need}, except with probability at most~$.1$. We conclude that $\Tilde{\bigo}\left(d_j^2 2^{-j}/\eps^2\right)$ copies of~$\rho$ suffice to distinguish $\rho = \sigma$ from $\|\rho[j,j] - \sigma[j,j]\|_1 > .05\eps/m^2$ with probability at least $.9 - .1 = .8$.  Repeating the test $\bigo(\log(m/\delta)) = \Tilde{\bigo}(1) \cdot \log(1/\delta)$ times increases the confidence to $1 - .5\delta/m$, as needed to complete the proof.
\end{proof}

\begin{lemma}
    \label{lem:upper-case-4}
    Suppose Case~2 of \Cref{lem:upper-cases} does not hold. Then, using entangled measurements, $\Tilde{\bigo}\left(\|\sigma^*\|_{1/2}/\eps^2\right)\cdot \log(1/\delta)$ copies suffice to test, with probability at least $1-\delta$, whether $\rho = \sigma$ or whether Case~3 holds.
\end{lemma}
\begin{proof}
    Fix distinct $j, j^\prime \in \mc{J}$ and assume, without loss of generality, that $d_j \geq d_j^\prime$. We will first test whether $\rho = \sigma$ or $\|\rho[j,j^\prime]\|_1 \geq .2\epsilon/m^2$ with confidence $1-\delta/m^2$, and then sum over all  $< m^2$ pairs. Now, let $\rho^\prime_{j,j^\prime}, \sigma^\prime_{j,j^\prime}$ denote the operators obtained by zeroing out all entries of $\rho,\sigma$ outside the principal submatrices indexed by $S_j \cup S_{j^\prime}$. Recall that $\rho[j,j^\prime]$ refers to the matrix obtained by zeroing out the two diagonal blocks of $\rho^\prime_{j,j^\prime}$, and similarly for $\sigma$. Denote $\hat{\rho}^\prime_{j,j^\prime} \triangleq \rho^\prime_{j,j^\prime}/ \tr(\rho^\prime_{j,j^\prime}),$ and $\hat{\sigma}^\prime_{j,j^\prime} \triangleq \sigma^\prime_{j,j^\prime}/ \tr({\sigma}^\prime_{j,j^\prime})$.  
    Recall that if Case~2 does not hold, then $\|\rho[j,j] - \sigma[j,j]\|_1$ and $\|\rho[j^\prime,j^\prime] - \sigma[j^\prime,j^\prime]\|_1$ are both at most $.05\epsilon/m^2$.

    First, consider the case when $\tr(\sigma^\prime_{j,j^\prime}) < .1\epsilon/m^2$. We will show that in this case, $\|\rho[j,j^\prime]\|_1 < .2\epsilon/m^2$; i.e., Case~3 does not hold. Assume, for the sake of contradiction, that $\|\rho[j,j^\prime]\|_1 \geq .2\epsilon/m^2$. Along with \Cref{lem:psd-block-1norm}, this implies that $\tr(\rho^\prime_{j,j^\prime}) \geq .2\epsilon/m^2$. Let $M \triangleq \rho[j,j] + \rho[j^\prime,j^\prime] - \sigma[j,j] - \sigma[j^\prime,j^\prime]$. First, we upper bound $\|M\|_1$:
    \begin{equation}
    \label{eq:case-4-subcase-1-contradiction}
        \|M\|_1 \leq \|\rho[j,j] - \sigma[j,j]\|_1 + \|\rho[j^\prime,j^\prime] - \sigma[j^\prime,j^\prime]\|_1 < .1\epsilon/m^2,
    \end{equation}
    where we use the triangle inequality followed by the lemma's hypothesis (Case~2 does not hold). Now, we can use our assumption to lower bound $\|M\|_1$:
    \begin{equation}
        \|M\|_1 \geq \|\rho[j,j] + \rho[j^\prime,j^\prime]\|_1 - \|\sigma[j,j] + \sigma[j^\prime,j^\prime]\|_1 = \tr(\rho^\prime_{j,j^\prime}) - \tr(\sigma^\prime_{j,j^\prime}) \geq .2\epsilon/m^2 - .1\epsilon/m^2 = .1\epsilon/m^2, 
    \end{equation}
    where in the first step we use the triangle inequality, and in the second step we use the fact that $\rho[j,j], \rho[j^\prime, j^\prime]$ form the two diagonal blocks of $\rho^\prime_{j,j^\prime}$, which is positive semidefinite (and similarly for $\sigma$). This contradicts \eqref{eq:case-4-subcase-1-contradiction}, thus showing that $\|\rho[j,j^\prime]\|_1 < .2\epsilon/m^2$ whenever $\tr(\sigma^\prime_{j,j^\prime}) < .1\epsilon/m^2$.

    Now, we consider the case when $\tr(\sigma^\prime_{j,j^\prime}) \geq .1\epsilon/m^2$. Let $\Pi_{j, j^\prime}$ be the projector onto the indices in $S_j \cup S_{j^\prime}$. First, we estimate $\tr(\rho^\prime_{j,j^\prime})$ to within additive error $.025\epsilon/m^2$ with confidence $1 - .5\delta/m^2$. This can be done by applying the POVM $\{\Pi_{j,j^\prime}, \mathbb{I} - \Pi_{j,j^\prime}\}$ individually to $\bigo(m^4 /\epsilon^2) \cdot \log(m^2/\delta) = \Tilde{\bigo}(1/\eps^2) \cdot \log(1/\delta)$ copies of $\rho$. If the estimate is more than $.025\epsilon/m^2$ far from $\tr(\sigma^\prime_{j,j^\prime})$, then we are confident that, $\rho \neq \sigma$ and can reject. Otherwise, we can assume
    \begin{equation}
    \label{eq:case4-subcase2-cond-1}
    \left|\tr(\rho_{j,j^\prime}^\prime) - \tr(\sigma^\prime_{j,j^\prime})\right| \leq .05\epsilon/m^2.
    \end{equation}
    Further, as $\tr(\sigma^\prime_{j,j^\prime}) \geq .1\epsilon/m^2$, this implies
    \begin{equation}
    \label{eq:case4-subcase2-cond-2}
        \tr(\rho_{j,j^\prime}^\prime) \geq \frac{1}{2} \tr(\sigma^\prime_{j,j^\prime}) = \Omega(d_j 2^{-j} + d_{j^\prime}2^{-j^\prime}).
    \end{equation}
    Next, we wish to lower bound $\|\hat{\rho}^\prime_{j,j^\prime} - \hat{\sigma}^\prime_{j,j^\prime}\|_{\HS}$.
    First, we rewrite the difference between the two normalized operators.
    \begin{equation}
        \hat{\rho}^\prime_{j,j^\prime} - \hat{\sigma}^\prime_{j,j^\prime} =  \begin{pmatrix}
            A & B \\ B^\dag & C
        \end{pmatrix},
    \end{equation}
    for some matrices $A \in \mbb{C}^{d_j \times d_j}, B \in \mbb{C}^{d_j \times d_{j^\prime}}, C \in \mbb{C}^{(d - d_j - d_j^\prime) \times (d - d_j - d_j^\prime)}$. As the off-diagonal blocks of $\hat{\sigma}^\prime_{j,j^\prime}$ are all 0, we have
    \begin{equation}
        \frac{\rho[j,j^\prime]}{\tr(\rho^\prime_{j,j^\prime})} =  \begin{pmatrix}
            \mathbf{0} & B \\ B^\dag & \mathbf{0}
        \end{pmatrix}.
    \end{equation}
    Thus, $2 \|B\|_1 = \|\rho[j,j^\prime]\|_1/\tr(\rho^\prime_{j,j^\prime})$. Now we are ready to bound $\|\hat{\rho}^\prime_{j,j^\prime} - \hat{\sigma}^\prime_{j,j^\prime}\|_{\HS}$:
    \begin{multline}
        \|\hat{\rho}^\prime_{j,j^\prime} - \hat{\sigma}^\prime_{j,j^\prime}\|_{\HS} = \left\|\begin{pmatrix}
            A & B \\ B^\dag & C
        \end{pmatrix}\right\|_{\HS}
        \geq \left\|\begin{pmatrix}
            \mathbf{0} & B \\ B^\dag & \mathbf{0}
        \end{pmatrix}\right\|_{\HS}
        = 2\|B\|_{\HS}
        \\ \geq \frac{2\|B\|_1}{\sqrt{d_j^\prime}}
        = \frac{\|\rho[j,j^\prime]\|_1}{\tr(\rho^\prime_{j,j^\prime}) \sqrt{d_j^\prime}}
         = \Omega\left(
            \frac{\|\rho[j,j^\prime]\|_1}{\tr(\sigma^\prime_{j,j^\prime}) \sqrt{d_j^\prime}}
        \right),
    \end{multline}
    where in the first inequality we use the fact that zeroing out entries of a matrix can only decrease its Hilbert--Schmidt norm, and in the second inequality we use the fact that $B$ has dimension $d_j \times d_j^\prime$ and $d_j^\prime \leq d_j$. Thus, it suffices to obtain a test for distinguishing with confidence $.9$ (which we will later increase to $1-.5\delta/m^2$) between $\rho = \sigma$ and $\|\hat{\rho}^\prime_{j,j^\prime} - \hat{\sigma}^\prime_{j,j^\prime}\|_{\HS} \geq \epsilon^\prime$, where 
    \begin{equation}
        \label{eq:upper-case4-epsilonprime}
        \epsilon^\prime = \Omega \left(\frac{\epsilon}{m^2 \sqrt{d_j^\prime} (d_j 2^{-j} + d_j 2^{-j^\prime})}\right).
    \end{equation}
    From \Cref{lem:hs-state-cert}, we see that this can be done using $\bigo(d_j^\prime (d_j 2^{-j} + d_{j^\prime}2^{-j^\prime})^2 m^4/\epsilon^2)$ copies of $\hat{\rho}^\prime_{j,j^\prime}$. Similarly to the proof of \Cref{lem:upper-case-3}, we only have access to copies of~$\rho$, not~$\hat{\rho}^\prime_{j,j^\prime}$. We handle this by applying the POVM $\{\Pi_{j,j^\prime}, \mathbb{I} - \Pi_{j,j^\prime}\}$ individually to 
    \begin{equation}
        \bigo \left(
            \frac{d_j^\prime (d_j 2^{-j} + d_{j^\prime}2^{-j^\prime})^2 m^4 }{\tr(\rho^\prime_{j,j^\prime}) \epsilon^2}
        \right)
        = \bigo \left(
            \frac{d_j^\prime (d_j 2^{-j} + d_{j^\prime}2^{-j^\prime}) m^4 }{\epsilon^2}
        \right)
    \end{equation}
    copies of $\rho$, where we use \Cref{eq:case4-subcase2-cond-2}. Except with probability .1, this gives us sufficiently many copies for the test from \Cref{lem:hs-state-cert}. Thus, we see that $\Tilde{\bigo}(d_j^\prime (d_j 2^{-j} + d_{j^\prime}2^{-j^\prime})/\eps^2)$ copies of $\rho$ suffice to test whether $\rho = \sigma$ or $\|\hat{\rho}^\prime_{j,j^\prime} - \hat{\sigma}^\prime_{j,j^\prime}\|_{\HS} \geq \epsilon^\prime$ with probability at least $.9-.1 = .8$. This can be improved to the desired confidence parameter $.5\delta/m^2$ by repeating the test $\bigo(\mathrm{log} (m^2/\delta))$ times. 
    
    Altogether, we have shown that $\Tilde{\bigo}(d_j^\prime (d_j2^{-j} + d_{j^\prime}2^{-j^\prime}) /\epsilon^2)\cdot \log(1/\delta)$ copies of $\rho$ suffice to test whether $\rho = \sigma$ or $\|\rho[j,j^\prime]\|_1 \geq .2\epsilon/m^2$ for some fixed, distinct $j,j^\prime \in \mc{J}$ with $d_j \geq d_{j^\prime}$, with the desired confidence $1-\delta/m^2$. Now, we will simply sum over all such pairs.
    \begin{align}
        \sum_{j \neq j^\prime, d_j \geq d_j^\prime} d_j^\prime(d_j2^{-j} + d_{j^\prime}2^{-j^\prime}) &\leq \sum_{j \neq  j^\prime, d_j \geq d_{j^\prime}} d_j^2 2^{-j} + \sum_{j \neq j^\prime,  d_j \geq d_{j^\prime}} d_{j^\prime}^2 2^{-j^\prime}
        \leq 2m \sum_{j \in \mc{J}} d_j^2 2^{-j}
        \\
        &\leq 2m \left(
            \sum_{j \in \mc{J}} d_j 2^{-j/2}
        \right)^2
        \leq 4m \left(
        \sum_{i \in [d]\backslash\stail} \sqrt{\lambda_i} 
        \right)^2
        = \bigo(m \|\sigma^*\|_{1/2}),
    \end{align}
    where in the first step we use $d_j^\prime \leq d_j$, in the penultimate step we use the fact that $\lambda_i \geq 2^{-j(i) - 1}$, and the final step follows from \Cref{def:bucketing-upper}. Thus, we see that when Case~2 does not hold, $\Tilde{\bigo}(\|\sigma^*\|_{1/2}/\epsilon^2)\cdot \log(1/\delta)$ copies suffice to test whether $\rho = \sigma$ or whether Case~3 holds, completing the proof.
\end{proof}

\bibliographystyle{alpha}
\bibliography{biblio}
\end{document}